%% file: main.tex
\documentclass[sn-mathphys,Numbered]{sn-jnl}

\usepackage{graphicx}%
\usepackage{multirow}%
\usepackage{amsmath,amssymb,amsfonts}%
\usepackage{amsthm}%
\usepackage{mathrsfs}%
\usepackage[title]{appendix}%
\usepackage{xcolor}%
\usepackage{textcomp}%
\usepackage{manyfoot}%
\usepackage{booktabs}%
\usepackage{algorithm}%
\usepackage{algorithmicx}%
\usepackage{algpseudocode}%
\usepackage{listings}%
\makeatletter
\@ifundefined{orcidlogo}{}{\let\orcidlogo\relax}
\makeatother
\usepackage{orcidlink}
\usepackage{orcidlink}

\usepackage{dirtytalk} 
\usepackage[acronyms,shortcuts,section=section,style=super,nopostdot,nogroupskip,nomain,nonumberlist]{glossaries} 
\glsdisablehyper 

\usepackage{cleveref}

\newacronym{PLM}{PLM}{Product Lifecycle Management}
\newacronym{MBSE}{MBSE}{Model-Based Systems Engineering}
\newacronym{DTI}{DTI}{Digital Twin Instance}
\newacronym{DTP}{DTP}{Digital Twin Prototype}
\newacronym{DTE}{DTE}{Digital Twin Environment}
\newacronym{DT}{DT}{Digital Twin}
\newacronym{MandS}{M\&S}{Modelling and Simulation}
\newacronym{VandV}{V\&V}{Verification and Validation}

\begin{document}

\title[Reusing Model Validation Methods for the Continuous Validation of Digital Twins of Cyber-Physical Systems]{Reusing Model Validation Methods for the Continuous Validation of Digital Twins of Cyber-Physical Systems}



\author*[1, 2]{\fnm{Joost} \sur{Mertens}{\hypersetup{pdfborder={0 0 0}}\orcidlink{0000-0002-8148-5024}}}\email{joost.mertens@uantwerpen.be}

\author[1, 2]{\fnm{Joachim} \sur{Denil}{\hypersetup{pdfborder={0 0 0}}\orcidlink{0000-0002-4926-6737}}}\email{joachim.denil@uantwerpen.be}

\affil[1]{\orgdiv{Applied Engineering: Electronics and ICT}, \orgname{University of Antwerp}, \orgaddress{\street{Groenenborgerlaan 171}, \city{Antwerpen}, \postcode{2020}, \state{Antwerpen}, \country{Belgium}}}

\affil[2]{\orgdiv{Ansymo/Cosys-lab}, \orgname{Flanders Make@University of Antwerp}, \orgaddress{\street{Groenenborgerlaan 171}, \city{Antwerpen}, \postcode{2020}, \state{Antwerpen}, \country{Belgium}}}



\abstract{\input{text/a-abstract}}




\maketitle

\input{text/b-introduction}
\input{text/c-related}

\input{text/d-approach}

\input{text/e-demonstration}
\input{text/f-discussion}

\input{text/g-conclusion}
\input{text/h-appendix}

\bibliography{references}

\end{document}

%% file: text/a-abstract.tex
One of the challenges in twinned systems is ensuring the digital twin remains a valid representation of the system it twins. Depending on the type of twinning occurring, it is either trivial, such as in dashboarding/visualizations that mirror the system with real-time data, or challenging, in case the digital twin is a simulation model that reflects the behavior of a physical twinned system. The challenge in this latter case comes from the fact that in contrast to software systems, physical systems are not immutable once deployed, but instead they evolve through processes like maintenance, wear and tear or user error. It is therefore important to detect when changes occur in the physical system to evolve the twin alongside it. We employ and reuse validation techniques from model-based design for this goal. Model validation is one of the steps used to gain trust in the representativeness of a simulation model. In this work, we provide two contributions: (i) we provide a generic approach that, through the use of validation metrics, is able to detect anomalies in twinned systems, and (ii) we demonstrate these techniques with the help of an academic yet industrially relevant case study of a gantry crane such as found in ports. Treating anomalies also means correcting the error in the digital twin, which we do with a parameter estimation based on the historical data.

%% file: text/b-introduction.tex
\section{Introduction}
\label{sec:introduction}

One of the challenges in twinned systems is ensuring the digital twin remains a valid representation of the system it twins. In its most general form, a digital twin consists of a digital model representing a real-world system (not necessarily physical), data gathered during the operation of that real-world system, and a way of adjusting the model based on that data if deemed necessary~\cite{wright2020}. With a model at its core, having that model be representative of the real-world system is paramount for providing accurate services. However, in their systematic mapping study on software engineering for digital twins, ~\citet{Dalibor2022} find that there is a lack of research in this direction. They pose the question of how quality assurance for digital twins is supported, and their answer is, to quote~\citet{Dalibor2022}: \say{. . . the number of publications considering the online verification of Digital Twins with their counterparts is vanishingly low . . . This implies a need for further research regarding the quality assurance of Digital Twins . . . during the runtime of both systems. Especially, the fidelity of Digital Twins, i.e., the verification that these properly represent the twinned system, needs further investigation.}. Digital twins are applied in a range of domains such as manufacturing, healthcare, urban/city planning, maritime and shipping, aerospace, automotive~\cite{botin2022,semeraro2021}. Given the different regulations in all of these domains, e.g. regarding safety or data ownership and usage, it is unlikely that the fidelity question has a single unanimous solution. Nonetheless, given our personal expertise, we look to contribute to the answer for the class of engineered systems called Cyber-Physical Systems.

A Cyber-Physical System (CPS) is a system that integrates computational and physical processes. Using (networked) feedback loops, the computational and physical processes interact and affect each other. The computational process allows for control and monitoring of the physical process~\cite{Lee2008}. Examples of such systems can be found in the automotive (traffic control, vehicles), avionics, and robotics industries. A way to deal with the complexities of Cyber-Physical Systems is to employ Model-Based Systems Engineering with \ac{MandS} in the system design. This entails modeling the physical process, e.g. as a  set of Ordinary Differential Equations (ODEs), and testing various control processes against this model.
In \ac{MandS}, through the process of \ac{VandV}, and specifically the validation part, modelers aim to prove that the model is representative of reality~\cite{Sargent1984,Schlesinger1979}. With this background in mind, it is clear that CPS and their accompanying models are prime contenders to be evolved into digital twins.

With the digital twin components and the CPS context in mind, we see that, at least during an engineered system's design time, the quality assurance problem posed by~~\citet{Dalibor2022} adds up to a \ac{VandV} process. Typically, in a \ac{MandS} life-cycle, the model is deemed ready for experimentation after the \ac{VandV} process\cite{balci2012life}. Despite normally being used at design time, this does not mean that the \ac{VandV} techniques cannot be used at runtime, and provide an answer to the last part of the proposed problem. 

In this paper, we present an approach that enables the reuse of model validation methods to continuously detect if a model deviates from the real-world system, also referred to as continuous validation in literature. The aim of this approach is to enable the evolution of a digital twin alongside its real-world counterpart. Our contributions are twofold: (i) we provide a generic approach that, through the use of validation metrics, is able to detect anomalies in twinned systems, (ii) we demonstrate these techniques with the help of an academic yet industrially relevant case study of a gantry crane such as found in ports. The remainder of this paper is structured as follows: in \cref{sec:background}, we discuss some background on the topics of digital twins and model validity. Continuing on, in \cref{sec:related-work} we discuss related work on the topics of continuous validation for digital twins and anomaly detection. Then, in \cref{sec:approach} we discuss our approach in more detail. Afterward, \cref{sec:case-study} describes our case study and evaluation. Next, \cref{sec:discussion} provides a discussion and future work. Lastly, \cref{sec:conclusion} wraps up the paper with a conclusion.


%% file: text/c-related.tex
\section{Background}
\label{sec:background}

In this background section we look at how a digital twin might be defined, what model validity concerns, and what validation metrics are.
\subsection{Defining the Digital Twin}
\label{sec:what-is-a-dt}

An historic overview of the term  digital twin shows that the definitions are numerous~\cite{Singh2021} and though they share common elements, no unanimous definition exists.

In our view, a digital twin is an amalgamation of various models and datasets, representing aspects (parts/subsets) of a real-world system, also referred to as the ``twinned system''. Over time, these models and data evolve along with that system. The data evolves by the mere act of collecting it, whereas the models are evolved either manually or automatically such that their behavior remains similar to that of the system's aspect they represent. Our view aligns closely with the INCOSE Systems Engineering Vision 2035~\cite{INCOSE2021}, which says \say{A Digital Twin is an approximation of reality that integrates many engineering models \ldots}. It also aligns with \citet{wright2020}, who succinctly state that the minimal components of a digital twin are a model of the twinned system, an evolving set of data from the twinned system, and a way of updating the model according to that data if needed. \citet{wright2020} focus on what sets apart a digital twin from a regular computerized model. We have shown this distinction in~\cref{fig:minimal-digital-twin}, where it can be seen that the digital twin is an instantiation of the base model that reflects one instantiation of the real-world system. This figure just aims to show the distinction, it does not show all other possible dataflows. Indeed, as the digital twin remains a model, data can be collected from it, just as it can be used in the replaying of scenarios, what-if analyses and predictions of the future. Furthermore, more advanced digital twin services such as optimizations of the twinned system or predictive maintenance also remain possible.

We do not expect a digital twin to be a monolithic model. In contrast to definitions such as \cite{Singh2021}'s, we also do not demand real-time state mimicking of the real-world system through the exchange of real-time data. In our opinion, saying a digital twin acts on the physical system is more appropriate than requiring perfect mimicry,  an idea also found in~\cite{Modoni2019}. Finally, the twin must also not use the best available models as required by \citet{Glaessgen2012}, rather the models must be appropriate for the goals the digital twin must fulfill and the services it offers.


Lastly, we cover two other commonly referenced definitions, which we will use later in the evaluation section (in addition to Wright and Davidson's\cite{wright2020} definition). They are Grieves and Vickers's~\cite{Grieves2017} and Glaessgen and Stargel's~\cite{Glaessgen2012} definitions. 

\citet{Grieves2017} give the following definition: \say{\ldots the Digital Twin is a set of virtual information constructs that fully describes a potential or actual physical manufactured product from the micro atomic level to the macro geometrical level. At its optimum, any information that could be obtained from inspecting a physically manufactured product can be obtained from its Digital Twin.}. The definition stems from \ac{PLM} and 
is rather open, as it speaks of virtual constructs, which can be interpreted as requirements models/documents, CAD models, behavioral models 
, historical runtime data, a bill of materials, etc. \citet{Grieves2017} also confirm this, stating these virtual constructs may (not must) be included in the digital twin, and the digital twin is not limited to those specific virtual constructs. Less open is the use of ``fully describes'', which means the digital twin of a system ought to describe every aspect of that system. In that regard, we think many digital twins are only partial digital twins, which is why in our view we opt to specify only an aspect of the twinned system can be represented.

\citet{Glaessgen2012} propose another definition: \say{A Digital Twin is an integrated multiphysics, multiscale, probabilistic simulation of an as-built vehicle or system that uses the best available physical models, sensor updates, fleet history, etc., to mirror the life of its corresponding flying twin. The Digital Twin is ultra-realistic and may consider one or more important and interdependent vehicle systems, including airframe, propulsion and energy storage, life support, avionics, thermal protection, etc.}. Their definition stems from an aeronautics and astronautics context, where current practices of vehicle certification, fleet management and sustainment based on heuristics and similitude will not suffice for future vehicles and missions~\cite{Glaessgen2012}. 
Compared to \citet{Grieves2017}'s definition, there is no more mention of a virtual construct. Instead, it is made specific to what kind of simulation is used.
The use of ``best available''  means there is some leniency to which data is considered acceptable, but we think the ``most appropriate'' rather than the ``best available'' models should be used.

\begin{figure}
    \centering
    \includegraphics[width = 0.75\textwidth]{figures/digital-twin-minimal.pdf}
    \caption{Visual representation of the difference between a (base) computerized model, and a minimal digital twin. The twin is updated throughout its lifecycle to accurately represent the real-world system, this is achieved through continuous data collection. In contrast, the base model does not evolve.}
    \label{fig:minimal-digital-twin}
\end{figure}

\subsection{Validity of Models}
\label{sec:validity-of-models}

The goal of \ac{MandS} is to utilize computational models to study reality. Every model is made with a set of response quantities in mind, that is, those properties that interest the modeler and for which they look to find answers or model predictions. However, a model remains an approximation of reality, and as such, care must be taken when interpreting its results. To review the representativeness of a model, a verification, validation and uncertainty quantification effort must be made, such that the model's limits are clearly understood~\cite{NAP2012}. Verification concerns itself with the solution accuracy of a computational model, validation concerns itself with the agreement between reality and computational models, and uncertainty quantification concerns itself with quantifying the effect of sources of error on the model's predictions. We continue on the topic of model validation. An early definition of validation can be found in~\cite{Fishman1968}, which states that \say{Validating a model means establishing that it resembles its actual system reasonably well}. Another definition, by the Society for Computer Simulation (SCS) states that validation is the \say{Substantiation that a computerized model within its domain of applicability possesses a satisfactory range of accuracy consistent with the intended application of the model}~\cite{Schlesinger1979}. This definition is carefully worded, and dissecting it yields the following insights:

\begin{enumerate}
    \item A model has a domain of applicability for which it is validated. If the model is to be used outside this domain, a new validation effort must be pursued.
    \item the model remains an approximation of reality, therefore the model's users must define a range of accuracy which they deem accurate enough for decision-making based on the model's results. To see if this range of accuracy is met, a comparison between the real world and the simulation is implied.
    \item a model has an intended purpose for which it is validated. If the model is to be used for another purpose, a new validation effort must be pursued. The purpose is closely linked to the properties of interest of the model.
\end{enumerate}

Other definitions of validation exist, e.g. by the Department of Defense and the American Institute of Aeronautics and Astronautics, for an overview of how those differ from the SCS definition, we refer to~\cite{Oberkampf2010}.

A more conceptual interpretation of validation is that of substitutability or interchangeability\cite{law2014simulation}. \citet{law2014simulation} states:\say{Conceptually, if a simulation model is `valid,' then it can be used to make decisions about the system similar to those that would be made if it were feasible and cost-effective to experiment with the system itself.}. Thus, there is this notion that if the decisions made through the simulation model are indistinguishable from those made with the real system, the model is valid.

The question of how to perform this validation effort differs between domains, and also differing philosophical views exist~\cite{beisbart2019computer}. This is unfortunate since the concept of digital twin is gaining traction in fields other than manufacturing~\cite{Dalibor2022}. Nonetheless, for engineered systems, in general those systems where simulation models are used extensively, we can resort to what \citet{Oberkampf2010} call “the restrictive notion of model validation”, that is, validating a model through comparison with experimental data. In this approach, a carefully crafted set of experiments are executed both on the real system, and on the to-be-validated model. The experiment describes for example the number of replications, the initial conditions and the stop times of the experiment. The comparison of the data happens through the study of properties of interest, also response quantities. The comparison of the response quantities happens through the use of validation metrics. These are metrics that generally calculate either a distance between data points or apply a statistical test to the data, and they may be preceded by data transformations such as transformations to different domains. \citet[Chapter~12.3]{Oberkampf2010} provide some recommended features for these validation metrics, a non-exhaustive list of metrics can be found in~\cite{Oberkampf2010,Maupin2017}. Common examples are the Root-Mean Square Error and Mahalanobis Distance. A final step in the validation effort is extrapolating the validation results if the experimental domain does not overlap the operational domain/domain of intended use.

Despite being normally applied at design time, there is an argument for reusing these model validation techniques at runtime. Indeed, most of the data necessary to apply the techniques also exists at runtime. However, a system in operation must perform its normal operation. Therefore, we cannot expect to have direct control over the real-world system to set up experiments as we please. At the very best, we might be able to perform experiments during scheduled downtime, though such downtime is normally strict and allocated for other purposes such as maintenance. As such, we assume we cannot perform experimental operations on the system and are limited to the data it generates in its normal operation. This brings along some limitations described below:

\begin{itemize}
    \item \textbf{Domain:} For the measured data, the data that can be collected is limited to the operational domain of the real system. For the simulated data, the entire valid domain of the digital twin can be covered. Without extrapolation of the measured data, the validation metrics can only be calculated over the union of those two domains, and are therefore limited to the operational domain of the real system.
    \item \textbf{Replications}: Replicated measurements of the real-world system rely on the system performing an identical operation multiple times in its normal operation. Furthermore, we lack control over the time between those replications. For the simulated data, replications are still possible at will.
\end{itemize}

In summary, it becomes harder to set up a carefully defined experimental design, but the model validation techniques remain applicable nonetheless.

\subsection{Validation Metrics}
\label{sec:validation-metrics}

The validation metrics we use in the implementation are described below. The mix of metrics gives an appropriate assortment of metrics that work with type 1 and type 2 data, that is, data where either no replications are done, or data where replications are present in one of the two datasets~\cite{Maupin2017}. Replications in the real world mean physically redoing the experiment, whilst replications in the simulation world mean simulating the model again. Because such simulations are generally deterministic, a replication relies on uncertain/stochastic model parameters for which the exact value is unknown, but for which we do have statistical information such as the distribution from which they can be sampled. A common example are initial conditions measured with a measuring device that has a measurement error. Replicating the simulation thus means drawing new samples for those model parameters. We opted for these metrics because, as we will see later, we can easily obtain type 1 and type 2 data, it is however harder to obtain type 3 data (data where both datasets are treated as uncertain). For more details on the metrics, we refer the interested reader to the cited references. By no means is this an exhaustive list of metrics, other metrics such as the Bayesian Validation Metric~\cite{Vanslette2019}, the probability density function area metric~\cite{Ferson2008,Roy2011}, the reliability metric~\cite{Rebba2008} and many more also exist and could be used.

To keep the notations consistent, the metrics are described as a set of experimental data $D$ and model predictions $P$, where a subscript $i$ indicates the $i$th element in the set. For $P$ and $D$ there also exists a vector $x$ containing the input variables associated with each prediction or datapoint. $x$ is likely to be time. When replicated simulations are used, we signify the sample mean over each of those predictions by $\overline{P}$. This does imply that the indexes of those samples overlap. If this is not the case, the data needs to be resampled. For the standard deviation of the prediction we use $\sigma_{P,i}$, when it is not known but multiple replications have been made, the sample standard deviation can be used instead.

\subsubsection{Root Mean Square Error}

The Root Mean Square Error between experimental data $D$ and predictions $P$ is given by\cite{Maupin2017}:

\begin{equation}
    d_{rmse} = \sqrt{\frac{1}{N}\sum_i{(\overline{P}_i - D_i)^2}}
\end{equation}

This thus calculates an average error per point and retains the original unit.

\subsubsection{Normalized Euclidean Distance}

The normalized Euclidean distance\cite{Maupin2017} is calculated as:

\begin{equation}
    d_i = \frac{\lvert P_i - D_i \rvert}{\sigma_{P,i}}
\end{equation}

When $\sigma_{P,i}$ is not known, but multiple replications of the simulation have been performed, we can calculate the sample standard deviation and use the sample mean $\overline{P}_i$ instead of $P_i$. Care must be taken that the standard deviation is not close to zero, otherwise those data points will skew the metric. We may consider excluding them if they pose a problem. We can reduce these distances to a scalar by taking the mean, or by calculating the total normalized Euclidean distance\cite{Maupin2017}, which is given by:

\begin{equation}
    d_ne = \sqrt{\sum_i{d_i^2}}
\end{equation}

The normalized Euclidean distance is unitless. When we exclude samples with a standard deviation of zero from the calculation, simply summing the distance will not allow for proper comparison between traces that exclude a different number of samples. One solution is to sum, then take the mean of the included samples as shown below. When we use the total normalized Euclidean distance in this paper, we refer to \cref{eq:tned}.

\begin{equation}
    d_ne = \sqrt{\frac{1}{N}\sum_i{d_i^2}}
    \label{eq:tned}
\end{equation}





\subsubsection{Average Relative Error}

The average relative error metric is\cite{Oberkampf2010,Oberkampf2006}:

\begin{equation}
    \left\lvert \frac{\widetilde{E}}{\overline{P}} \right\rvert_{avg} = \frac{1}{(x_i - x_0)}\int_{x_0}^{x_i} \left \lvert \frac{D_i - \overline{P}_i}{\overline{P}_i}  \right \rvert \,dx\
\end{equation}

Where $\widetilde{E}$ represents the estimated error, and $x_0$ and $x_i$ represent the first and last input variable, e.g. time. The average relative error metric is unitless or in percent. Care must be taken when the prediction mean is close to zero, otherwise those data points will skew the metric. In our calculations, we do exclude them.




\subsubsection{Maximum Relative Error}

The maximum relative error is given by:\cite{Oberkampf2010,Oberkampf2006}:

\begin{equation}
    \left\lvert \frac{\widetilde{E}}{\overline{P}} \right\rvert_{max} = \max_{x_0 \leq x \leq x_i} \left \lvert \frac{D_i - \overline{P}_i}{\overline{P}_i}  \right \rvert
\end{equation}

The same remark about the mean close to zero holds here as well.

\section{Related Work}
\label{sec:related-work}

We find related work in the domains of online/continuous validation of digital twins, and anomaly/fault detection. This related work is discussed in more detail in the following subsections.

\subsection{Online/Continuous Validation of Digital Twins}

\citet{Overbeck2021} describes the continuous validation of a digital twin of a production system. In this process, they iteratively validate and model-update their digital twin. In their validation, they use two validation metrics that are calculated on the Key Performance Indicator (KPI) ``number of produced parts'': the relative variation and the normalized Root Mean Square Error. The digital twin they apply this to is that of a production system of car parts, consisting of two production areas connected to each other with conveyors. If the thresholds that are set for the metrics are surpassed, the models are updated with appropriate new parameters.

In a similar vein, \citet{Lugaresi2022} describes the online validation of digital twins in a manufacturing system by applying the dynamic time warping methods, which is a time-series analysis that assess the distance between two sequences. They further split up the threshold setting into a warning and a failure threshold. The technique is applied on the digital twin of a small lab-scale manufacturing facility consisting of two workers and two conveyors. In~\cite{Lugaresi2023} they extend this approach with an additional technique from bio-informatics: that of the Longest Common Sub-sequence (LCSS) and the Modified Longest Common Sub-sequence (mLCSS).

Though they do not mention continuous validation, \citet{Munoz2022} presents a method that measures the similarity between a digital twin and its real-world counterpart through trace alignment. Their approach assumes the traces might not be aligned in time and that samples of the traces might be missing. To combat this assumption, they use the Needlemann-Wunsch algorithm to align the traces. Only then do they calculate a distance metric on the samples, in this case, the Fréchet distance based on Euclidean and Manhattan distance.

Conceptually, the related work is closely related to the techniques presented in this paper, especially the work of ~\citet{Overbeck2021} and \citet{Lugaresi2022}. All the listed techniques look to detect when a system is diverging from the initial behavior by utilizing a digital twin. There are however some differentiating points between our techniques and the related work. Firstly, the utilized simulation formalism in the digital twin differs. ~\citet{Overbeck2021} and~\citet{Lugaresi2022} focus on production lines or manufacturing systems, which are modelled with worker nodes and conveyors in a discrete-event simulation. Such models are generally characterized by distributions on the processing time of the worker nodes and conveyors, i.e. by statistical distributions. In contrast, the models in CPS are physics-based. In our example, we use a kinematics model solved by a Runge-Kutta method. Secondly, and a result of the different simulation formalisms, are the properties of interest of the simulation, their timescale and also the applicable validation metrics. ~\citet{Lugaresi2022} use dynamic time warping, which is a measure of similarity between two traces traditionally used for classification. Dynamic time warping is robust w.r.t. time shifts, as it allows for the mapping of samples of one trace to one or more samples of the other trace (not necessarily at the same index). For a production line, where the amount of products produces is of more interest than the exact time they are produced, such a characteristic is quite useful. \citet{Munoz2022} use a similar metric as well. For physics-based simulations the properties of interest are the states of the system over time, here we prefer to have metrics that work on samples with the same index, since not only the shape of the state-trace is of importance, but also the transient behavior. We do not want a metric that corrects for time shifts in that behavior, which is why we opt for more traditional validation metrics. \citet{Overbeck2021} use more traditional metrics such as the normalized root-mean-square error, but they use KPIs of the production line on a day-by-day basis. This is a completely different timeframe, as we will be comparing traces with samples separated sub-seconds apart.


\subsection{Anomaly/Fault Detection}

Assuring the quality of a digital twin can also be interpreted as detecting the occurrence of anomalies/faults, if we assume all is well with the twin until such an anomaly is detected. In case of such a detection, further action must be taken. Fault detection and diagnosis has been around for a long time, and methods can be broadly classified as model-free or model-based methods \cite{gertler1998}. Recall that the goal of fault detection is slightly different from digital twin quality assurance, in that in the former we wish to make sure the digital models are representative, whereas in the latter we want to detect faults in the real system. As such, model-free fault detection methods such as physical redundancy or special sensors \cite[Chapter~1]{gertler1998} are not particularly useful for the digital twin cause. Other techniques such as limit checking \cite[Chapter~1]{gertler1998}, data stream analysis~\cite{Schmidt2020} or log analysis~\cite{Harada2017} are, but as they are lacking a model they are less relevant than the model-based methods. Model-based methods then, utilize an explicit model of the monitored system, and rely on the concept of analytical redundancy~\cite{gertler1998}. Similar to how in physical redundancy parallel sensor measurements are compared, in analytical redundancy the model computations are compared with the system measurements. The differences, also called residuals, are evaluated and can indicate faults in the system. To generate the residuals, techniques such as Kalman filtering and parameter estimation can be applied. These detection techniques can be applied to continuous systems~\cite{gertler1998}, but also hybrid~\cite{Narasimhan2007,Zhao2005} systems. It is clear that this approach shows similarities with the model validation approach, only the view is different. In the validation we look to prove the model is valid, whereas in the fault detection we look to detect a fault in the system with the model, invalidating the system.

\citet{Feng2021} look into the application of these anomaly detection techniques in the engineering of digital twins. They successfully applied a Kalman Filter-based anomaly detection in an incubator system, where the anomaly that can be detected is the inadvertent opening of the lid of the incubator.

Nowadays, the focus has shifted to not only detecting faults, but also detecting security intrusions, and learning-based approaches are being used in this field~\citet{Luo2021}. A thorough overview of existing work on deep learning-based anomaly detection can be found in~\cite{Luo2021}. We also see that the digital twin concept is gaining traction in this field, as ~\citet{Xu2021} describe a technique to detect anomalies based on Generative Adversarial Networks (GAN) and a system's digital twin with historic and runtime data. The digital twin data is specifically used to compute ground truth labels used in the training of the GAN. Their anomaly detector showed increased performance compared to state-of-the-art anomaly detectors in three benchmark datasets.

%% file: text/d-approach.tex
\section{Approach}
\label{sec:approach}

Regular model validation compares simulation and real-world experiments to define if a model is valid. We adapt this concept of model validation to the run-time validation of the system and its twin. For this, we need to (a) be able to log the data from the real-world system, (b) define what a real-world experiment entails during the operation of the system, (c) be able to set up and run the virtual experiment(s) and (d) calculate appropriate validation metrics. Our framework consists of an architecture with physical and computerized components and an accompanying workflow. This workflow is strongly inspired by general model validation workflows such as those in~\cite{Oberkampf2010}. By repetitively applying this workflow during the system's operation, we achieve what is called, in literature, a continuous or continual validation of the system. 

\subsection{Architecture}
\begin{figure}
    \centering
    \includegraphics[width = 0.7\textwidth]{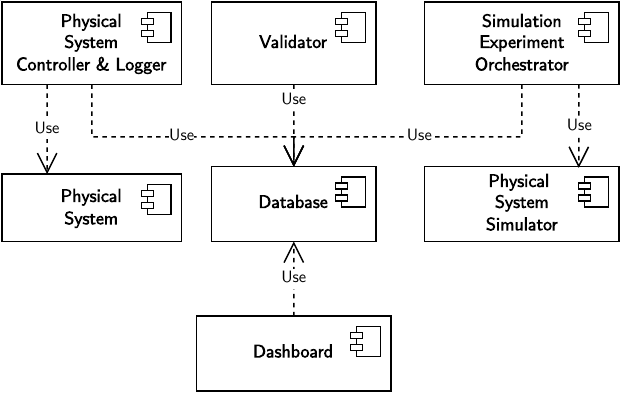}
    \caption{General architecture needed to enact the workflow.}
    \label{fig:component_diagram}
\end{figure}

The general architecture of the framework is shown in the component diagram in \cref{fig:component_diagram}. This diagram contains the minimal subset of components needed to enact the workflow but may be expanded upon in the implementation (as discussed in ~\cref{sec:case-study}). The component diagram is kept as general as possible. Therefore, we do not expose interfaces on the components but adhere to more straightforward usage relationships and do not include specific subcomponents. In what follows, we briefly discuss the components.

\begin{itemize}
    \item The \textbf{Physical System} is the system operating in the real world. It exposes two interfaces to the controller: one for control signals and one for data logging. We assume that internally the \textbf{Physical System} handles its low-level control.
    \item The \textbf{Physical System Controller \& Logger} provides high-level control of the \textbf{Physical System}, and handles the task of logging both the sent control signals as well as the measured data to a \textbf{Database} component. In contrast with low-level control, the high-level control is on a different level of abstraction, it is all the control that is needed such that the \textbf{Physical System} can enact the control that is provided by a digital twin service. For example in general robotics, the high-level control may track a specific trajectory generated by a digital model. The low-level control handles the specific actuation of the robot's motors to attain that goal. Whether a controller must be implemented depends on the interface provided by the \textbf{Physical System}. In case of what~\citet{Kritzinger2018} calls a digital shadow, only the logging component must be implemented.
    \item The \textbf{Physical System Simulator} is a computerized model with simulation capabilities representing the \textbf{Physical System}. Internally, this component consists of separate simulations for different aspects of the system. This component does not need to be complete, that is, not every aspect of the \textbf{Physical System} must be simulated, with the caveat that non-modelled aspects cannot be simulated, and hence also not validated.
    \item The \textbf{Simulation Experiment Orchestrator} is in control of instantiating the \textbf{System Simulator}, and logging their data to the \textbf{Database}.
    \item The \textbf{Validator} calculates validation metrics based on the data logged to the database by the \textbf{Controller} and the \textbf{Simulator}. It also writes away those results to the \textbf{Database} for further processing. Lastly, it is also tasked with detecting deviations and providing corrected system parameters to those components that need them, likely the \textbf{Controller}, \textbf{Simulator} and \textbf{Physical System Simulation}.
    \item The \textbf{Database} provides a storage facility for data coming from the other components.
    \item The \textbf{Dashboard} provides a visualization (dashboard) of the data contained in the database. This is completely optional, but a welcome addition for industrial relevance. Common choices are Grafana, Kibana and Tableau. These dashboards are not only limited to visualization but can also handle more advanced tasks, e.g., sending out email notifications on events.
\end{itemize}

A small remark is that \cref{fig:component_diagram} only demonstrates the components for a single physical system. This diagram can be expanded for multiple physical systems, in which case there will be multiple pairs of \textbf{Physical System} and \textbf{Physical System Simulator} components. Whether or not the \textbf{Controller}, \textbf{Validator} and \textbf{Simulation Experiment Orchestrator} components are also multiplied is left up to the specific implementation. A likely course of action is to multiply the  \textbf{Controller} and keep a single \textbf{Validator} and \textbf{Simulation Experiment Orchestrator}.

\subsection{Workflow}

\begin{figure}
    \centering
    \includegraphics[width = \textwidth]{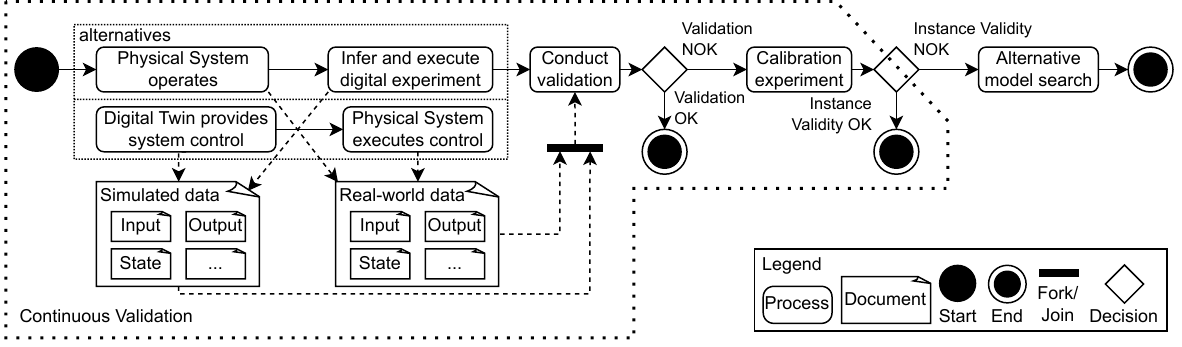}
    \caption{Visualization of the workflow. The continuous validation workflow does not include the alternative model search step, as indicated by the dotted line.}
    \label{fig:continuous_validation_workflow}
\end{figure}

The accompanying workflow is shown in \cref{fig:continuous_validation_workflow} as an activity diagram, with the addition of documents to show how data is used. The initial ideation of this workflow stems from~\cite{Mertens2021} and it was further refined in~\cite{Mertens2023}. The workflow shows the continuous validation of the system: continuously comparing real-world data with simulated data. In line with the terminology used in the domain of model validation, we refer to the process of obtaining the data as a real-world experiment and a digital experiment. The main difference with regular model validation is that rather than crafting a specific validation experiment and executing it on the real system and in simulation for comparison, we treat the data coming from the normal operation of the system as the real-world data for the validation experiment. Additionally, we also need to obtain simulated data to compare to. In the workflow, we distinguish two cases, the first dealing with a legacy system that does not use a digital twin service to provide system controls, and the second a new system which does.

In the first case, to obtain the simulated data, we must extract (infer) a validation experiment from the data provided by the regular operation of the system. The assumption here, is that the \textbf{Physical System} is operating normally and thus that the controller is logging real-world data to the database. To perform a validation, simulated data is also needed. This means, we extract the control data and use it to execute a virtual experiment (simulation) to create the simulated data.  It is the task of the  \textbf{Simulation Experiment Orchestrator} to do this extraction. 

\begin{figure}
    \centering
    \includegraphics[width = 0.6\textwidth]{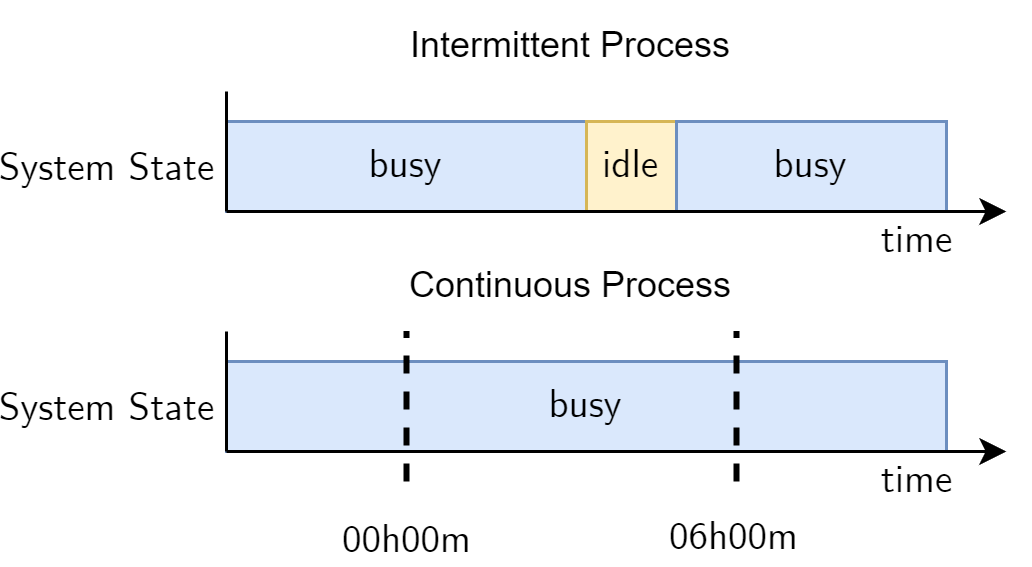}
    \caption{System state of an intermittent and continuous process over time.}
    \label{fig:experiment-delimiting}
\end{figure}

The inference of this experiment depends on how the \textbf{Physical System} executes, either intermittent (discontinuous) or continuously. In an intermittent process, there generally exists a routine that can be discerned from the system state. Either the system is performing this routine, or it is idle. A continuous process, on the other hand, does not have these distinct states, rather it only knows the busy state. This is shown in \cref{fig:experiment-delimiting}.

In the intermittent process, the most straightforward way to infer an experiment from the data is to treat the execution of a single routine as the experiment. The experiment delimiting can be done more elaborately by using multiple cycles of the routine, which might be needed if statistical measures are needed. In such case, care must be taken in selecting the number of cycles, since this will have a ``smoothing'' effect on measures such as the mean. It is possible that the routine cycle is of such length that it should be treated as a continuous process, which we discuss next.

In a continuous process, there is no clear start or stop of the routine cycle, instead, delimiting an experiment can be done by other means such as time-based (e.g. the last 5 minutes) or event-based (e.g. a property of interest is equal to a specific value). \Cref{fig:experiment-delimiting} shows a time-based delimiting from 0h to 6h. The same remarks regarding the selection of the number of experiments apply to the continuous process as well.

In the second case, when a digital twin of the system is present, the control of the system may be simulated in advance, after which the \textbf{Physical System} must only enact that control. For a routine-based system, this calculation may happen for example for each routine execution. For a continuously running system, a sliding window can be used, such that while the \textbf{Physical System} is enacting the current window, the next window can be calculated by the \textbf{Simulation Experiment Orchestrator}. It should be noted that in this alternative, the component diagram must be extended with a usage relationship between the \textbf{Controller} and the \textbf{Simulation Experiment Orchestrator}. This latter case is much more straightforward, since the simulated data merely follows from the regular operation of the system. As we will see in \cref{sec:case-study}, our case study fits this latter description.

After obtaining the two datasets, the \textbf{Validator} commences the calculation of the validation metrics. The set of calculable metrics depends on the provided datasets and the underlying assumptions on that data (e.g. assumption of normality). Through comparison with acceptable values of each of the metrics, the \textbf{Validator} determines the validity of the model.

When the model is deemed valid, no further steps are required. When it is deemed invalid,  we must try to evolve the digital twin such that it once again mirrors the real-world system. We distinguish two cases: when the system has structurally changed, and when it has not. Firstly, when the system has not structurally changed, that is, the governing equations remain the same, but only parameters have changed, mirroring the model can be done by calibrating the model. With a calibration experiment, a parameter estimation can be performed. This can be done by solving a minimization problem, e.g. minimizing the sum of squared errors (least squares) or other scalar cost functions. To do so, various tools exist, for example Scipy's~\cite{2020SciPy} \textit{Scipy.optimize} package with the \textit{scipy.optimize.minimize} function, Matlab's parameter estimator~\cite{matlabparameterestimator} or pyPESTO~\cite{schalte2023}. If such a parameter estimation manages to correct the validity of the model instance, the continuous validation process is finished. When it cannot correct the model validity, we find ourselves in the second case of model evolution, that is, the system has structurally changed. Consequently, we must search for an alternative model. This is a harder challenge, which we do not consider in the continuous validation. To solve it, we can resort to (hybrid) system identification techniques that can find a set of equations/models that appropriately describe the system~\cite{lauer2019,MORADVANDI2023}.

\subsection{Validation Metric Analysis}
\label{sec:metric-sensitivity-analysis}
One of the problems with the validation metrics mentioned in \cref{sec:validation-metrics}, is that they offer little intuition for which problems each metric would be good. The way the metric is calculated gives information about the type of data it can be used on, or if for example it puts more weight on large errors. Inspecting our metrics, we see that the normalized Euclidean distance grows large for standard deviations close to zero and that the average/maximum relative error are similarly impacted by prediction means close to zero. Gaining an intuitive notion of how the metric performs for the specific problem at hand remains hard. One of the ways to deal with this issue is to perform a study of the metrics under simulated conditions, e.g. such as done by \citet{liu2011}. With two models, a correct one and incorrect one, they inspect the model rejection percentage in function of the number of replicated measurements for type 3 data. An alternative approach, which we will take, is to inspect the behavior of the metric in function of the size of the error. Such an analysis inspects the sensitivity of the metric to the size of the fault, to which the metric might react linearly, quadratically, logarithmically, or perhaps it might not react at all. Such an analysis yields the intuitive notion of how the metric performs. If time and resources allow it, it should be performed in depth for every metric and every type of fault. The only caveat is that the analysis is limited to the faults we can foresee in the system.

%% file: text/e-demonstration.tex
\section{Case Study}
\label{sec:case-study}

Our case study concerns a digital twin system of a gantry crane. In the following subsections, we describe this system, the way it is instantiated, the evaluation objectives and procedure, and the evaluation results.

\subsection{Physical System}
\label{sec:physical-system}

\begin{figure}
    \centering
    \includegraphics[width =0.6\textwidth]{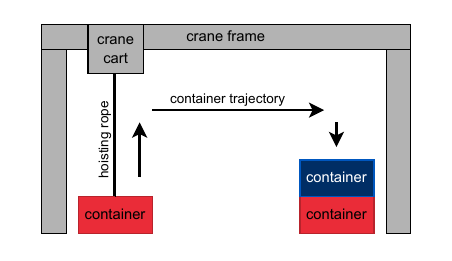}
    \caption{Abstract illustration of the routine of a gantry crane picking up and dropping off a container.}
    \label{fig:crane-routine}
\end{figure}

\begin{figure}
    \centering
    \includegraphics[width = \textwidth]{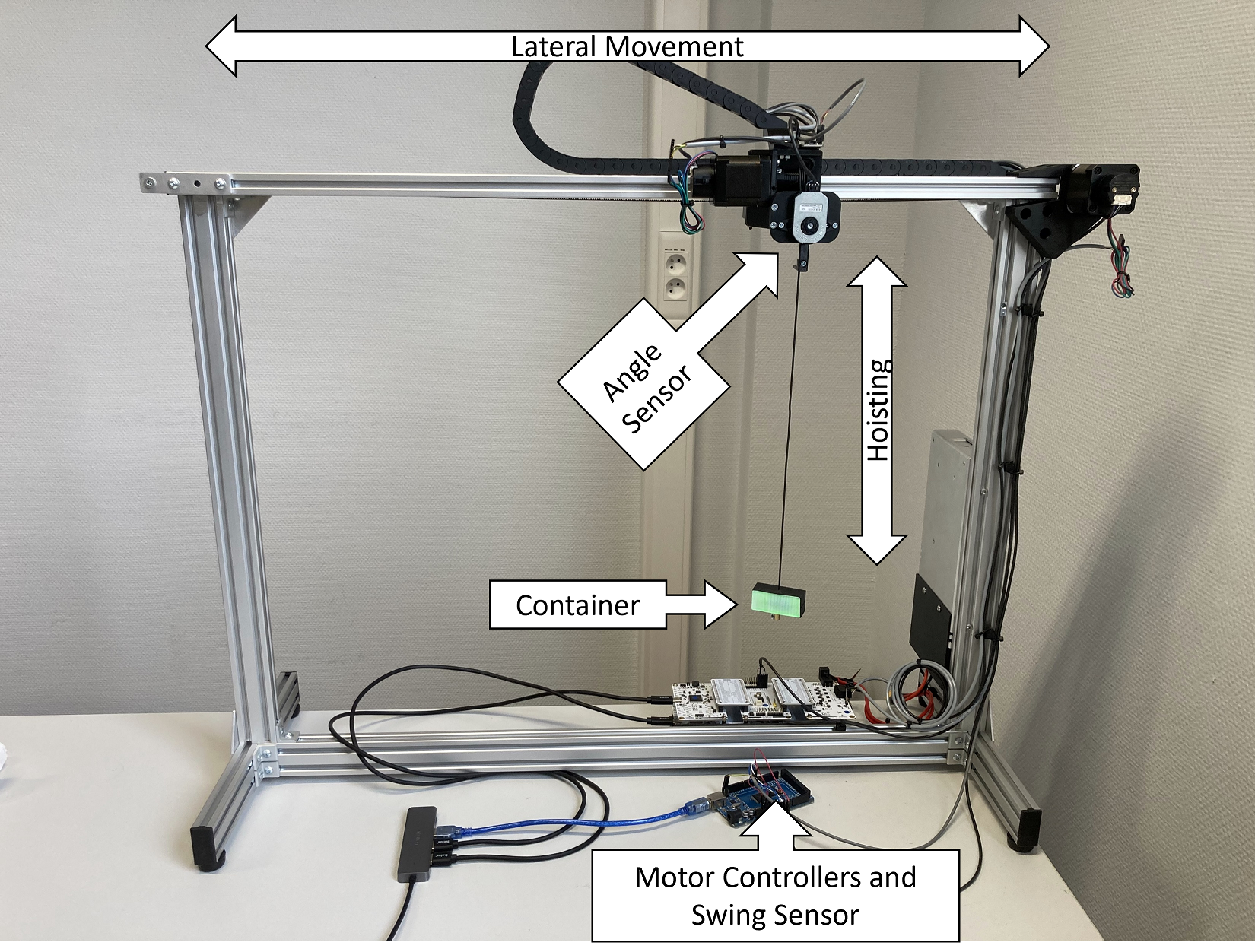}
    \caption{Photograph of the miniaturized gantry crane.}
    \label{fig:the-real-deal}
\end{figure}

Our case study is inspired by gantry cranes as seen in harbor terminals and consists of a miniaturized gantry crane. These cranes repetitively load and unload containers onto and from moored container ships, as is abstractly shown in \cref{fig:crane-routine}. This move requires some thought, since to move in an optimal way, the container should arrive at the lowering position with a minimal residual swinging motion. The real-world system that we are testing on is lab-scale miniaturized gantry crane, as shown in \cref{fig:the-real-deal}. This lab-scale crane is approximately scaled 1:10 to a real gantry crane, with a height and width of about 50 x 70 cm. It moves a container of 5 cm x 2 cm x 2 cm, which is approximately the scaled down equivalent of a 20ft standard container.

The miniaturized crane has two degrees of freedom: it can move containers horizontally, and it can hoist containers vertically. The motors and motor controllers driving this movement are two Trinamic motors (model Trinamic QSH4218-10000) and motor controllers (model Trinamic TMC4671). The motors have internal optical encoders and with the Field Oriented Control (FOC) provided by the controllers very precise movement is possible. The controllers handle the  entire FOC algorithm in hardware. For hoisting the container, a cart with a rope-drum attached straight to the motor axle is used. To measure the angle between the hoisting rope and the cart, an additional encoder (model CUI Devices AMT102-V) has been fitted with ``feeler needles'' as to minimally impact the container's swinging motion. 

For communication with a host computer, the motor drivers expose an SPI based interface through which they can be configured and controlled. For logging, they expose a UART interface that is accesible in parallel to the main SPI interface. This minimizes any impact of logging on the motor controllers' operation, since the logging does not share bandwidth with the control. To connect to this UART interface, we used a USB-UART adapter (model FTDI TTL-232R-RPI). The additional encoder measuring the angle is read by an Arduino Mega which communicates the measurements over the Arduino's USB-UART interface to the computer.

To move the gantry crane, the \textbf{Physical System Controller \& Logger} moves in three stages: hoisting up, performing a lateral move and hoisting back down. For hoisting/lowering, it simply sets the target height, for the lateral move, it must set the target motor position, and feed the controllers with their instantaneous maximum velocity to follow a specific trajectory. The movement of the container from the loading to the unloading position (or vice versa) is what we call the routine operation of this system. It shows discretized behavior, and we call it a ``run'' throughout the remainder of this paper. In this case, the controller connected to the system is a motion trajectory generator and follower written in Python. To perform the lateral move of the routine operation of the gantry crane, the controller performs two steps: (1) it generates an optimal trajectory given a start and end position and a set of constraints and (2) it makes the \textbf{Physical System} follow that trajectory while logging data during this enactment. To generate the trajectories, we utilized the Rockit toolbox~\cite{gillis2020} to define and solve an optimal control problem. In this control problem, we use the \textbf{Physical System Simulator's} model, which is an Ordinary Differential Equation (ODE) of the crane's kinematics, and its associated parameters and constraints. The \textbf{Controller} stores the generated trajectory, as well as the logged enactment of it, in the \textbf{Database}.

\subsection{Instantiation of Architecture}

\begin{figure}
    \centering
    \includegraphics[width = 0.8\textwidth]{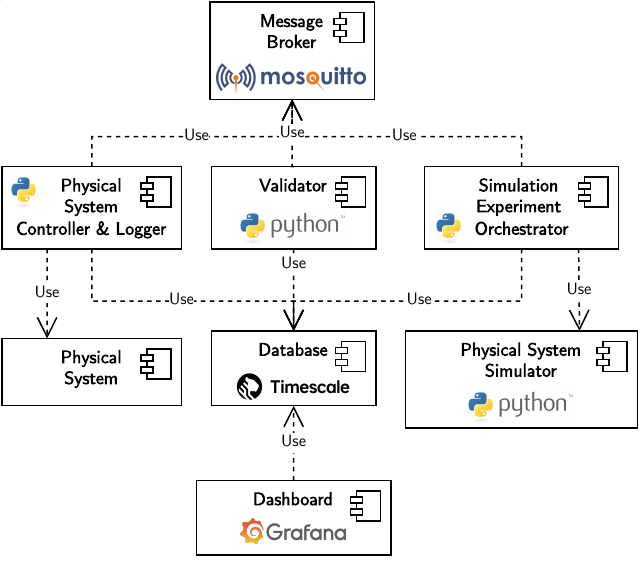}
    \caption{Architecture of the case study.}
    \label{fig:component-diagram-implemented}
\end{figure}

\Cref{fig:component-diagram-implemented} shows the component diagram of the use-case. The components from the general architecture in \cref{fig:component_diagram} have been labeled with their implementation, be it a general purpose programming language or a specific computer program. In the case of the physical system, this is not done, since it cannot be captured that way. Lastly, we note the addition of a \textbf{Message Broker} component that was not present in the original component diagram. Next we discuss the choices made for the various components, starting with the new component.

\begin{figure}
    \centering
    \includegraphics[width=\textwidth]{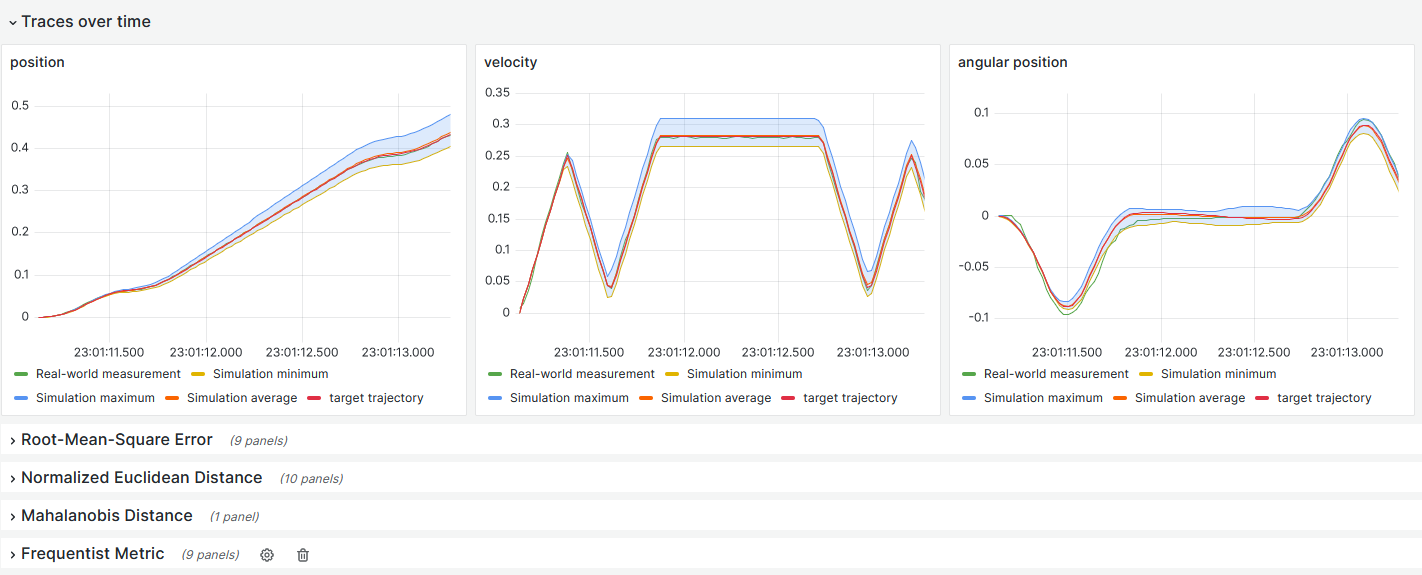}
    \caption{Grafana dashboard showing the optimal, measured and simulated traces of a run of the system. The validation metrics rows can be folded open for further investigation.}
    \label{fig:grafana-dashboard}
\end{figure}

\begin{itemize}
    \item The \textbf{Message Broker} is added to allow for alerting between the \textbf{Controller}, \textbf{Validator}, and \textbf{Simulation Experiment Orchestrator} components. This component is not necessarily needed, since the \textbf{Validator} and \textbf{Simulation Experiment Orchestrator} could poll the \textbf{Database} periodically for new data logged by the \textbf{Controller}, but it does allow them to react more quickly. For example, in the current implementation, the \textbf{Simulation Experiment Orchestrator} already starts simulating before the \textbf{Controller} has completed the trajectory. We opted for the open-source Eclipse Mosquitto~\cite{Light2017} as a broker and used Eclipse Paho to connect to the broker in the other components.
    \item The \textbf{Physical System} has previously been described in detail in \cref{sec:physical-system}.
    \item  The \textbf{Controller} has also previously been described in detail in \cref{sec:physical-system}. After storing the generated trajectory in the \textbf{Database}, the \textbf{Simulation Experiment Orchestrator} is notified through the \textbf{Message Broker} to start simulating with that trajectory as input. After storing the logged data, it signals the \textbf{Validator} through the \textbf{Message Broker} that new measured data is available.
    \item The \textbf{Physical System Simulator} is a python program incorporating a Runge-Kutta solver, which is used to solve the system's set of ODEs. This set of ODEs describe the kinematics of the crane. Associated with this simulator and simulation is a configuration file containing the parameters that characterize our specific \textbf{Physical System}. Running the simulation makes use of the ODE solvers in the Scipy~\cite{2020SciPy} library.
    \item The \textbf{Simulation Experiment Orchestrator} is also a Python program which sets up parallelized simulations based on the \textbf{Physical System Simulator}. It does so using Python's multiprocessing module. It also handles the sampling of stochastic variables for each replication.
    \item The \textbf{Validator} is also implemented in Python and calculates and stores the validation metrics covered previously in \cref{sec:validation-metrics}.
    \item We use Timescale as a \textbf{Database}. Timsescale is an extension of Postgres for time-series data. Without going into too much detail, the database is structured as follows: a machine's operation is stored in the measurement table. Each measurement has a quantity and an associated run. Each run corresponds to one routine operation. The ideal trajectory of each of those operations is stored in the trajectory table. Each of the runs can also be simulated, and a simulation replicated, so the simulationdatapoint table stores that information. Lastly, there are tables for each of the validation metrics (in fact more since some validation metrics can use 2 tables, one for local and one for global metrics). The complete entity relation diagram of the database, as well as a more detailed discussion, can be found in \cref{fig:ER-diagram} and \cref{sec:database-tables} of the appendices.
    \item For the \textbf{visualization}, we used Grafana as a dashboarding technology. Grafana enables us to easily visualize the system's operation and the calculated validation metrics. The dashboard is split up in rows, the first one being the measured and simulated trajectories, and all other rows contain the validation metrics for that trajectory. \Cref{fig:grafana-dashboard} shows these rows.
\end{itemize}

\subsection{Evaluation Objectives}

In the evaluation of our approach, we look to answer the following four  questions.

\textit{Q1: Is the implemented use case a digital twin?}

With this first question, we check whether the implementation fits in the three digital twin definitions as discussed previously in \cref{sec:what-is-a-dt}. The purpose is simple: given our proposed contributions, there is no point in evaluating further if our use case cannot be considered a digital twin.

\textit{Q2: How do the validation metrics behave?}

In this question, we look to explore the behavior of the validation metrics in function of the error size in a simulation study as proposed in \cref{sec:metric-sensitivity-analysis}. The purpose is to gain more insight in how the metrics perform for this case study.

\textit{Q3: Can our continuous validation approach show that the model stays valid during operation, and can it detect when it is no longer valid?}

This question looks to answer if we can detect when the real-world system diverges from the model prediction based on the model validation metrics. In other words, it is about the first decision diamond in \cref{fig:continuous_validation_workflow}. To answer this question, we devised a testing scenario that is described in more detail in the next section.

\textit{Q4: Can we evolve the digital twin, such that it again matches the real-world system?}

This last question involves the second decision diamond from \cref{fig:continuous_validation_workflow}. We check if, after detecting the change, we can correct the model such that its behavior is once again accurate?

The answers to questions 2 and 3 then form the final verdict on the main contribution of this paper: \textit{Can model validation methods be used to enable/support the evolution of digital twin models alongside their real-world system?}

\subsection{Evaluation Procedure}

To answer Q2, Q3 and Q4 we follow similar procedures, but Q2 is wholly in simulation, whereas for Q3 and Q4 we use the complete architecture as previously described. For all these questions, we need to observe the system under normal conditions to find appropriate thresholds, and under diverged conditions to see if the thresholds are sufficient to detect the diverged condition. For Q2, we will use the thresholds to inspect the sensitivity, whereas for Q3 and Q4, these thresholds are used in the validation OK/NOK decision diamond of the workflow in \cref{fig:continuous_validation_workflow}.

For Q2, we stick to simulating a single trajectory, which we replicate 50 times. We pick the threshold value as the maximum observed value in those 50 replications.
 We did so for the three quantities of position, velocity and angular position, because these are the quantities that are both simulated and directly measurable. This process of choosing the thresholds is automated, it just finds the maximum observed metric value under normal conditions.

To set the normal condition thresholds for Q3 and Q4, we let the crane operate normally for 10 separate trajectories. Similarly to Q2, we pick the threshold value as the maximum observed value in those 10 trajectories. We again did so for the three quantities of position, velocity and angular position. This process was again automated.

Next, we should define the diverged conditions. Preferably this would be a degraded condition in which the crane still produces traces that are relatively close to the normal traces, this way we can see if the threshold detection scheme is sensitive enough to detect minor divergences. Very abrupt divergences, such as the belt snapping or a motor driver burning out are therefore no options. Options such as an incorrect rope length due to the rope slipping on the drum or worn roller bearings of the cart are much more gradual and preferable. For Q2, we opted to go with the incorrect rope length, whereas for Q3 and Q4 We opted to go with worn roller bearings of the cart. An incorrect rope length affects the swinging frequency of the container, worn roller bearings make the movement of the cart more strenuous, which would reduce the acceleration of the cart given the motor's power, and consequently the top speed it can attain in a certain timespan.

For Q2, we performed simulations with incorrect rope lengths ranging from -10\% to +10\% of the target length in steps of 0.5\%, each of those replicated 30 times. Using the incorrect rope length also simplifies the comparison, only the angular position is affected by the incorrect rope length. This is sufficient to get an idea of the behavior of the validation metrics.

For Q3 and Q4, wearing out the roller bearings physically was not an option, instead we tried to emulate this effect by tightening the adjustment screws of the cart. However, in doing so, we found that it is hard to replicate exactly the same tightness on those setscrews, which makes it hard to reset them to exactly their original/normal state. To alleviate this issue, we decided to take an inverted approach, that is, rather than attempting to lower the achievable velocities physically and keeping the simulated velocity at the original level, we keep the cart's maximum physical velocity the same and  increase the maximum velocity of the simulation. The achieved effect is the same: the attained physical velocity is lower than the simulated maximum velocity. The advantage is that, since the physical system is not adjusted, going back to previous conditions only requires adjustments in software. The disadvantage is that for different maximum velocities, the shape of the generated trajectory changes.



To test the thresholds, we increased the maximal velocity of the motors in the model by respectively 2, 5, 10, 15 and 20\% above the normal value (0.281 m/s). Alongside that increase, the maximum acceleration was checked to ensure the increased velocity could be reached within the time constraints of the model. Due to the high acceleration of the motors, this required no change. For each of those increases, 10 runs were performed by the gantry crane. 

\subsection{Evaluation and Demonstration}

\subsubsection{Q1: Is the implemented use case a digital twin?}

To answer this question, we compare the setup of our use case to three digital twin definitions. We opt for the three definitions mentioned in \cref{sec:what-is-a-dt}, that is, \citet{wright2020}, \citet{Grieves2012}and \citet{Glaessgen2012}. We map the components of the architecture in \cref{fig:component-diagram-implemented} onto the elements in each of those definition.

\citet{wright2020} state there are three important parts of a digital twin: a \say{model of the object}, an \say{evolving set of data related to the object} and a \say{means of dynamically updating or adjusting the model in accordance with the data}.

\begin{itemize}
    \item The \textbf{model of the object} is the ODE model used in the \textbf{Physical System Simulator}, that same model is also used by the \textbf{Physical System Controller} in the calculation of the trajectories.
    \item The \textbf{evolving set of data related to the object} is stored in the \textbf{database}, and continuously supplemented by data from the \textbf{Controller}, \textbf{Validator} and \textbf{Simulation Experiment Orchestrator}.
    \item The \textbf{means of dynamically updating or adjusting the model in accordance with the data} is the process of updating the model parameters when the validation process discovers a mismatch. This remains to be proven by \textit{Q4}, but we can already tell that this is the case.
\end{itemize}

For Grieves's\cite{Grieves2012} definition, which can be found in \cref{sec:what-is-a-dt}, the elements map as follows:

\begin{itemize}
    \item The \textbf{virtual information construct} is the ODE model used in the \textbf{Physical System Simulator}, that same model is also used by the \textbf{Physical System Controller} in the calculation of the trajectories. In contrast to the definition, this \textbf{virtual information construct} does not \say{fully describe the physical system from the micro atomic level to the macro geometrical level}.
    \item The \textbf{potential or actual physical manufactured product} is the \textbf{Physical System}.
    \item The \textbf{information that could be obtained from inspecting a physically manufactured product} is the logged data stored in the \textbf{Database}. In contrast to the definition, we are not yet at the optimal point where \textbf{all} the data of the physical system is stored in that database.
\end{itemize}

For Glaessgen and Stargel's\cite{Glaessgen2012} definition, which can be found in \cref{sec:what-is-a-dt}, the elements map as follows:

\begin{itemize}
    \item The \textbf{probabilistic simulation} is the ODE model used in the \textbf{Physical System Simulator}, that same model is also used by the \textbf{Physical System Controller} in the calculation of the trajectories. It is probabilistic in the sense that we perform replications to propagate uncertainties of inputs/parameters through the simulation. In contrast to the definition, our simulation is not a \textbf{multiphysics} nor \textbf{multiscale} nor is it the \textbf{best available physical model}.
    \item The \textbf{as-built vehicle or system} is the \textbf{Physical System}.
    \item The \textbf{the best available, sensor updates, fleet history, etc.} are the data logged throughout the architecture and stored in the \textbf{Database}.
    \item When it comes to the movement of the crane, our model is realistic, though not \textbf{ultra-realistic}, since it is a linearization around the operating point.
    \item (Optional) our digital twin does not consider \textbf{interdependent vehicle systems}, e.g. in the case of the crane the electrical and actuation systems.
\end{itemize}

Summarizing, it seems that for Grieves's\cite{Grieves2012} and Glaessgen and Stargel's\cite{Glaessgen2012} definitions, our system is lacking some elements to fully match the definition. This mostly concerns completeness or the use of the best models available. For \citet{wright2020}'s definition, which is also formulated less strictly, all elements are present. These shortcomings in two of the definitions have us state that the answer to this question is \textit{no, the current setup does not adhere to the definitions closely enough to be considered a digital twin of the entire system, instead it is a digital twin of only an aspect/subset of that system}. We already gave some remarks to the strictness of Grieves's\cite{Grieves2012} and Glaessgen and Stargel's\cite{Glaessgen2012} definitions in \cref{sec:what-is-a-dt}, we just add that for us, being considered a digital twin of an aspect of the system is sufficient to be considered a digital twin of the system, and is also sufficient for the remaining research questions.

\subsubsection{Q2: How do the validation metrics behave?}

\begin{figure}
    \centering
    \includegraphics[width=\textwidth]{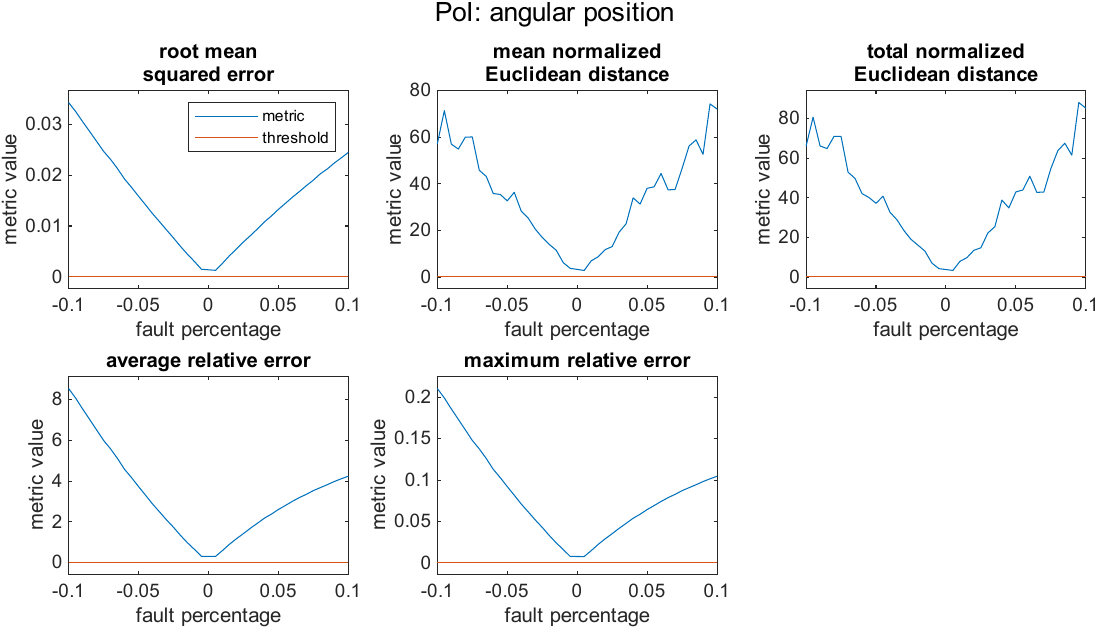}
    \caption{Behavior of the metrics for the angular position property of interest.}
    \label{fig:Q2-metrics}
\end{figure}

The behaviors of the various metrics are shown in \cref{fig:Q2-metrics}. In essence, they all exhibit the same behavior: as the fault increases, be it positively or negatively, the metric value also increases. From the shape of the metric value, we can tell that linearly increasing the fault has a nonlinear effect on the metric. This makes sense, since in a simple pendulum, the swing period is a function of the square root of the rope length. The simulations are very close to the reference trajectory, hence, the threshold is near zero. A final note is on the straight line between -0.05\% and +0.05\%, this line is only there due to the plot linearly connecting the data points. If we had included the 0\% fault, the data point would end up in a V on the threshold line. What remains to be seen from this study, is if these metrics carry over to the real world, where measurements are generally noisier, and if they carry over to different faults. As an answer to this question, \textit{we now do know how the metrics are likely to behave}.

\subsubsection{Q3: Can our continuous validation approach show that the model stays valid during operation, and can it detect when it is no longer valid?}

\begin{figure}
    \centering
    \includegraphics[width=\textwidth]{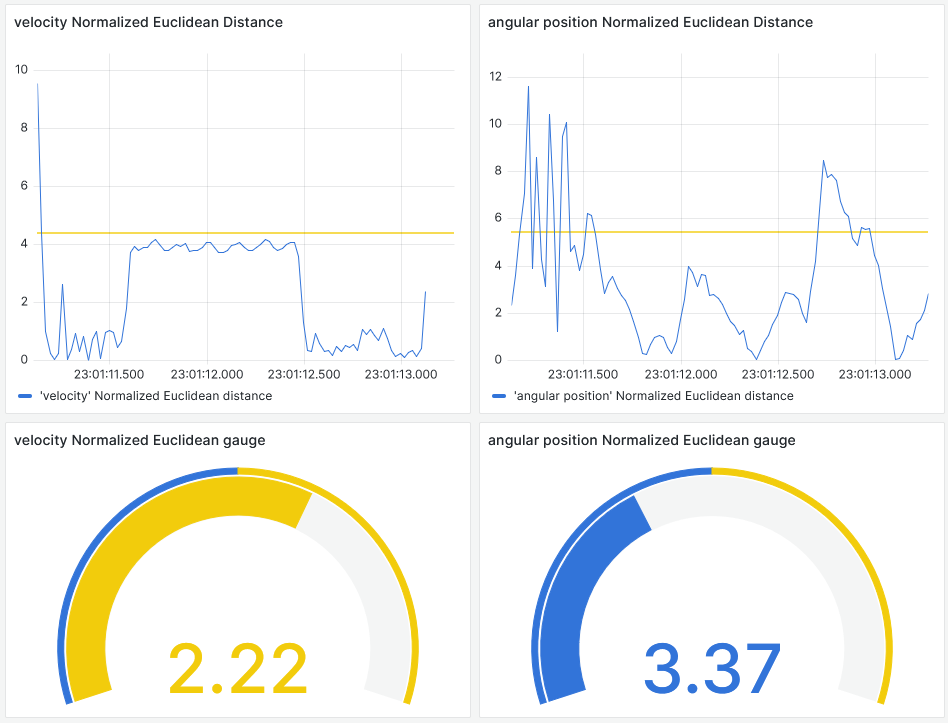}
    \caption{Threshold gauges in the dashboard for the normalized Euclidean distance. The dashboard contains more graphs and gauges, but only two are shown for brevity.}
    \label{fig:grafana-dashboard-thresholds}
\end{figure}

\begin{figure}
    \centering
    \includegraphics[width=\textwidth]{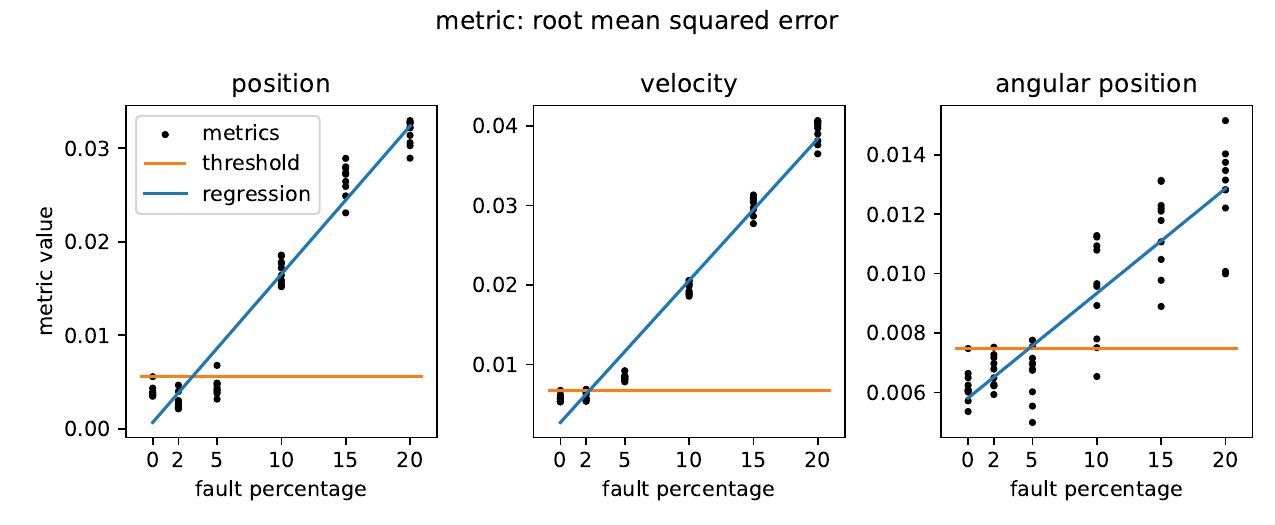}
    \caption{Root mean squared error for position, velocity and angular position in function of the fault percentage.}
    \label{fig:results-rmse}
\end{figure}

\begin{figure}
    \centering
    \includegraphics[width=\textwidth]{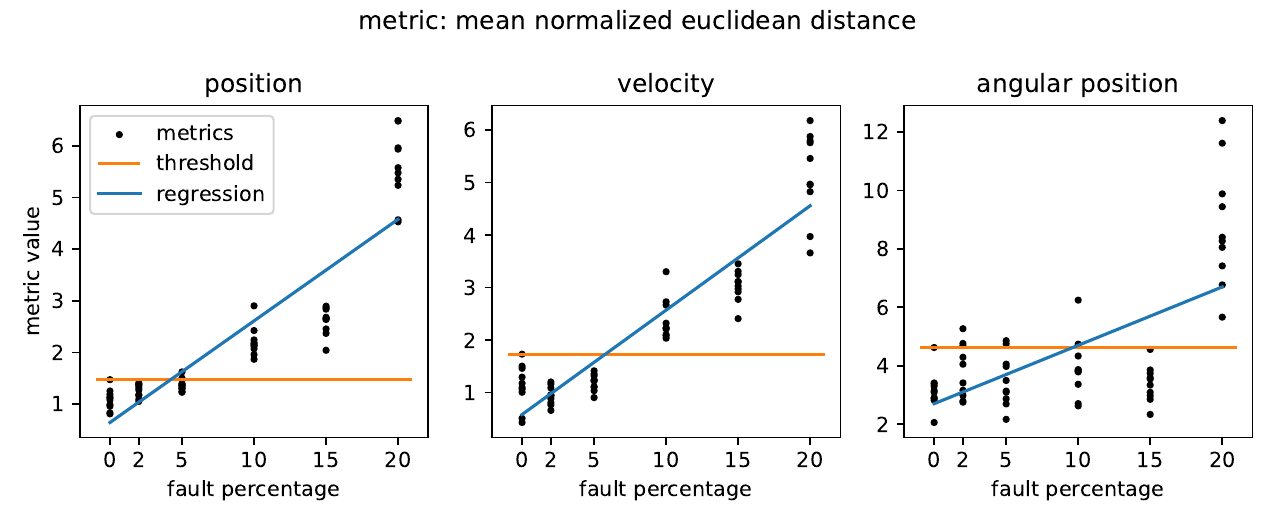}
    \caption{Mean normalized Euclidean distance for position, velocity and angular position in function of the fault percentage.}
    \label{fig:results-meanned}
\end{figure}

\begin{figure}
    \centering
    \includegraphics[width=\textwidth]{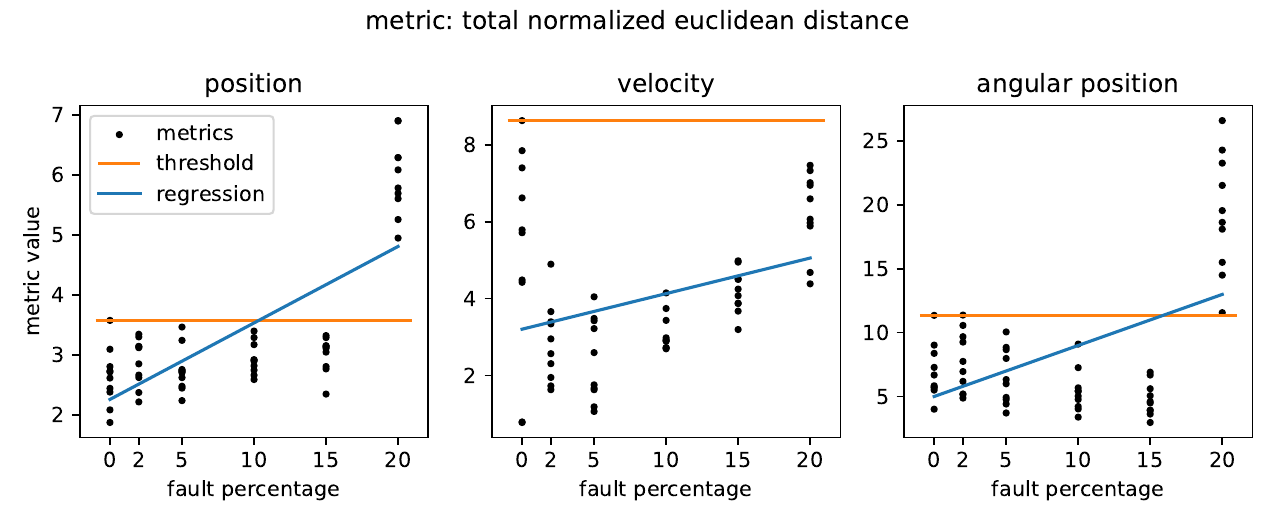}
    \caption{Total normalized Euclidean distance for position, velocity and angular position in function of the fault percentage.}
    \label{fig:results-totalned}
\end{figure}\

\begin{figure}
    \centering
    \includegraphics[width=\textwidth]{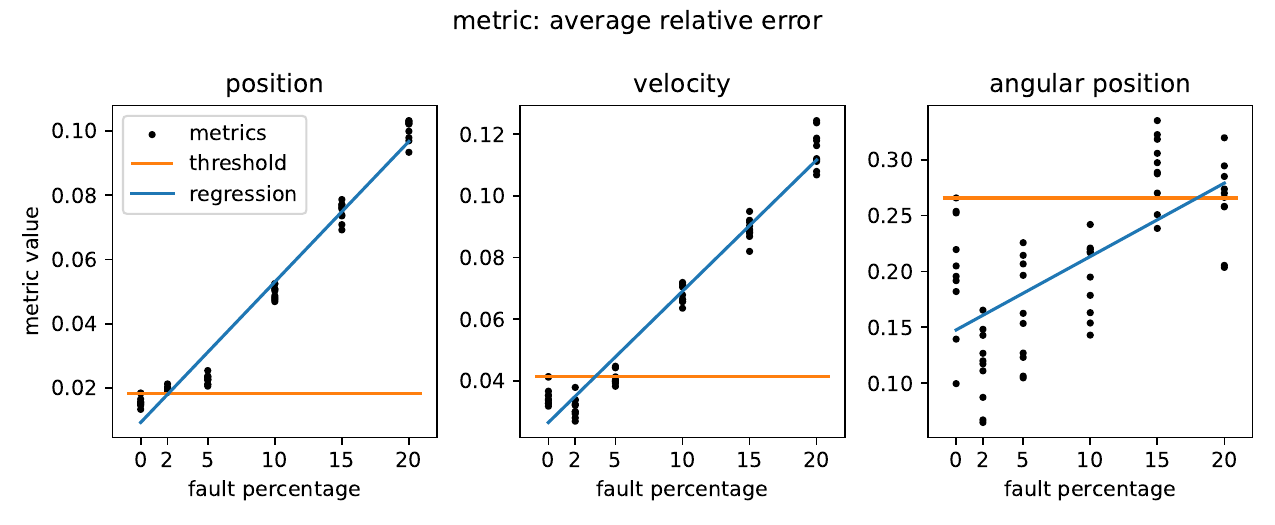}
    \caption{Average relative error for position, velocity and angular position in function of the fault percentage.}
    \label{fig:results-avg-rel-err}
\end{figure}

\begin{figure}
    \centering
    \includegraphics[width=\textwidth]{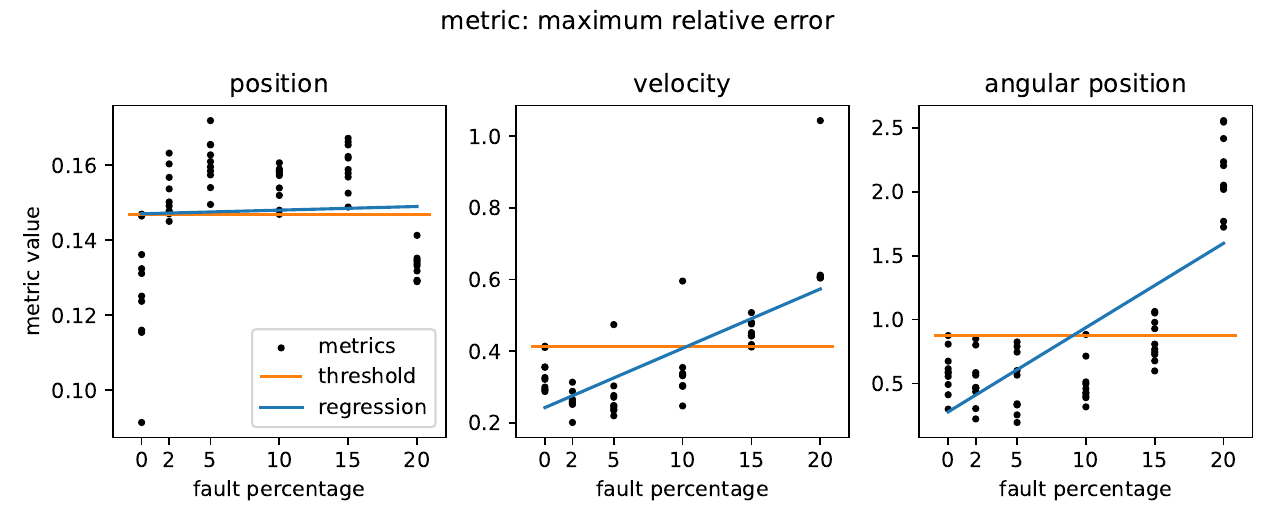}
    \caption{Maximum relative error for position, velocity and angular position in function of the fault percentage.}
    \label{fig:results-max-rel-err}
\end{figure}

The results are shown per metric in figures in \cref{fig:results-rmse,fig:results-meanned,fig:results-totalned,fig:results-avg-rel-err,fig:results-max-rel-err}. In the dashboard, the thresholds are visualized with gauges, which trigger to a warning color once the threshold is passed, an example is shown in \cref{fig:grafana-dashboard-thresholds} for the normalized Euclidean distance of a run.

We briefly discuss each of the metrics:
\begin{itemize}
    \item The root-mean-square error shows a clear linear trend upward as the fault increases in size. For the position, the fault is detected from 10\% and upward, for the velocity from 5\% and upward and for the angular position from 15\% and upward. Especially at the lower fault percentages, no detections occur, because the metric value does not breach the threshold.
    \item The mean normalized Euclidean distance shows a similar trend, crossing the threshold for faults of 10\% and larger. For the angular position, the metric value hovers around the threshold and seems unable to properly detect faults of less than 20\%. Because the regression crosses the threshold at a higher fault percentage, it seems less sensitive than the root mean squared error.
    \item The total normalized Euclidean distance is the first metric to not perform as expected. We see values that hover around or below the threshold, even for larger fault percentages.
    \item The average relative error fares well for the cart position and cart velocity. For the position quantity it detects faults from just 2\% and for the velocity it starts detecting them at 5\% and upward. For the angular position, it fails to have proper detection.
    \item The maximum relative error does not work for the position property. For the velocity and the angular position, it does show an increasing trend in metrics passing the threshold as the fault percentage increases, but only for larger fault percentages.
\end{itemize}

These results show that only some of the metrics are appropriate for the detection of  the roller bearing type of fault, despite the  simulation study indicating all metrics were fit for the rope-length fault. The ones that are fit do manage to detect the fault from an acceptable error percentage and onwards. To improve the detections or make them more robust, one could combine quantities and/or metrics in a majority vote detection scheme. 

Thus, to answer the question: \textit{yes, the approach can detect when the real-world system cannot attain the simulated trajectory. It can also detect that we are operating under normal conditions from a certain fault percentage onwards.}

\subsubsection{Q4: Can we evolve the digital twin, such that it again matches/mirrors the real-world system?}

\begin{figure}
    \centering
    \includegraphics[width=0.95\textwidth]{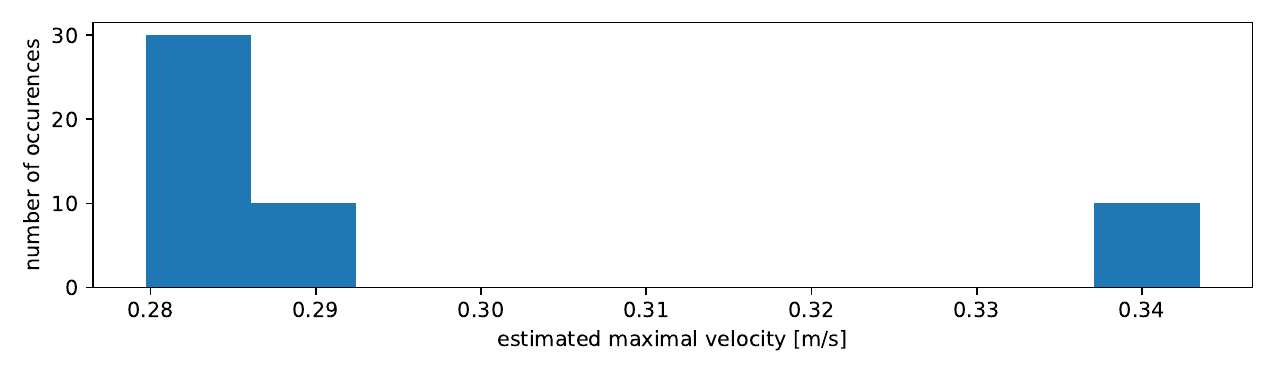}
    \caption{Histogram for the estimated maximal velocity parameter with the maximal velocity from the target trajectory as initialization condition.}
    \label{fig:maximal_velocity_histogram_max_init.png}
\end{figure}

\begin{figure}
    \centering
    \includegraphics[width=0.95\textwidth]{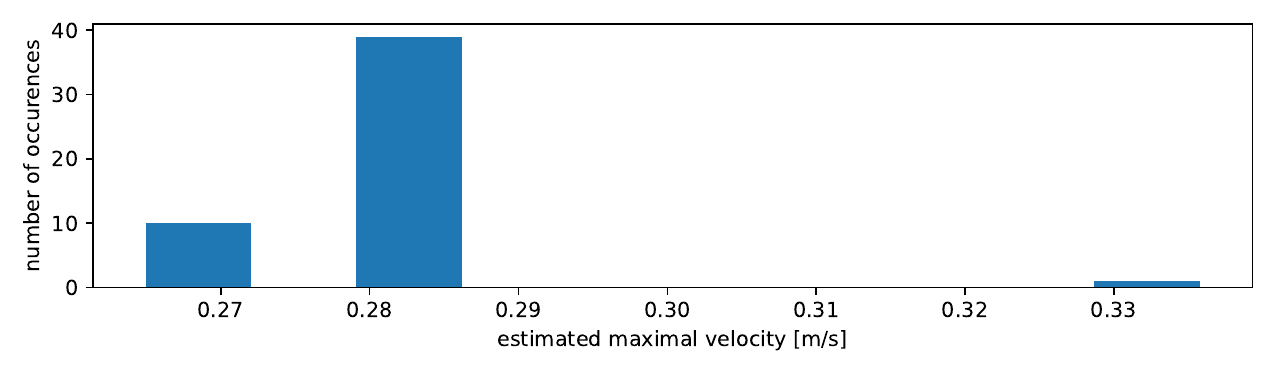}
    \caption{Histogram for the estimated maximal velocity parameter with 90\% of the maximal velocity from the target trajectory as initialization condition.}
    \label{fig:maximal_velocity_histogram_90_max_init.png}
\end{figure}

We use \textit{scipy.optimize.minimize} to estimate the correct parameters of the model, specifically the changed maximum motor velocity. We do so by minimizing the sum of squared errors for the position and velocity traces using the Nelder-Mead method. For reference, we extract the average maximum measured velocity and the accompanying standard deviation for the 50 divergent traces.
This is the target value the parameter estimation should attain if it worked perfectly. The average of the maxima is 0.282 m/s, with a standard deviation of 0.0003 m/s. We can then perform the parameter estimation, normally this would be done specifically on the trace that breached the threshold. To show the variability in the parameter estimation results, we calculate the estimation for each of the 50 divergent traces. The parameter estimation gives good, but not perfect results, and is highly dependent on the initial guess of the parameter. A good initial guess is the maximal velocity of the reference trajectory, since the system was able to attain this value in the past. Doing so for our 50 divergent traces, we find a mean of 0.294 m/s with a standard deviation of 0.024 m/s, on average an overestimation of the real parameter by 2.4\%. The histogram of this estimation is shown in \cref{fig:maximal_velocity_histogram_max_init.png}. Rather than fully automating the estimation, we could also gain knowledge from inspecting the trace, and see that the maximal velocity is lower than anticipated. If we use that knowledge to set the initial value in the parameter estimation to 90\% of the maximal velocity of the reference trajectory, we get a mean of 0.279 m/s with a standard deviation of 0.01 m/s, which is very close  to the reference and on average underestimates the parameter by 1.1\%. The accompanying histogram is shown in \cref{fig:maximal_velocity_histogram_90_max_init.png}. Underestimating this parameter is in this case preferable over overestimation. Alternatively, we could even use the maximal velocity from the measurement as the initialization condition, though if we are doing this we might also just resort to querying the maximum measured value directly from the database and skip the parameter estimation altogether.
Thus, the answer to this question is \textit{yes, within a tolerance, the twin can be evolved again to match the real-world system given proper initial conditions of the parameter estimation}.

%% file: text/f-discussion.tex
\section{Discussion \& Future}
\label{sec:discussion}

In what follows, we cover some points of discussion concerning the results, and propose future work.

\subsection{Discussion}

There remain some points of discussion, both with the current results and hypothetical situations not shown in the current results, into which we dive deeper.

\begin{itemize}
    \item \textbf{Applicability of metrics:} We saw in the results that not all metrics are suited for threshold-based detection. Some metrics show change, but only for large increases of the maximum velocity and are therefore too insensitive. Others seemingly show no change for changing conditions. This despite the earlier simulation study showing good results for all metrics. It seems hard to predict which metrics will be applicable and which will not from a simulation study.

    A likely culprit for this effect could be our approach of increasing the actual-simulated velocity delta virtually, rather than physically degrading the roller bearings. This is because increasing the velocity virtually also makes the controller generate a different optimal trajectory. This trajectory might be executed slightly differently, requiring a different threshold. Given the discretizable nature of the system, discovering such thresholds for each of the possible trajectories is a feasible solution to this problem.
    \item \textbf{Effect of small standard deviations:} Some of the metrics calculate a sample standard deviation and/or variance for further calculation. However, when this value is 0, the metric resolves to \textit{inf} or \textit{NaN}. When replicating simulations without at least slightly randomizing the initial conditions, it is guaranteed that the samples at $t=0$ will yield a standard deviation of 0. Furthermore, time steps close to $t=0$ likely also have a sufficiently small standard deviation to highly impact the metric, which will then tend to either very large or very small numbers. Furthermore, when dealing with a quantity that shows some oscillatory behavior, such as the angular position or angular velocity of the container on the gantry crane, there will also be points of convergence in the execution trace, where a standard deviation of nearly 0 is also observed, again causing the \textit{inf} or \textit{NaN} result. We observed this firsthand and alleviated it with success in the normalized Euclidean distance results by excluding samples with a sufficiently small standard deviation from the calculation. 
    \item \textbf{Threshold setting:} The classic problem from model validation of deciding on an acceptance threshold for the metrics also returns here. There is no ready-made answer for deciding on a threshold; in this case, we simply picked the maximum observed value during normal operation as the threshold. If we wanted a more sensitive threshold at the cost of false positives, we could allow for a certain percentage of type I error under normal conditions. This means a false positive rate of at most a certain percentage for the hypothesis that the model has diverged. Such a threshold makes the detection scheme more sensitive, but will also incur unneeded parameter estimations due to the false positives. 
    
    Another thought is that, technically speaking, when the models in the digital twin have been appropriately validated during the design process, we should already have these values from the design process. We can reuse them here, but this requires an extensive \ac{VandV} effort.
    \item \textbf{Gantry crane controller robustness:} The gantry crane controller is, given its simplistic nature, not robust to interference from the outside, for example, gusts of wind that move the container. That is to say, we could detect something wrong, but we cannot pinpoint it to that gust of wind. In a lab environment, where there are no such influencing factors, this is not an issue. Were we to use the crane in an outside environment, a more robust control scheme is needed to handle such interference. This issue also deals with the validity of the crane's model, which, we might say, is valid in an indoors lab environment and not in an outside harbor environment.
    \item \textbf{Generality of the setup:} Though we used an architecture that seemed reasonably general to us, we only provide one implementation of it. As such, we cannot claim that the architecture and workflow are generally applicable. Likely, every setup will need some ad-hoc elements, such as the message broker in our case. Nonetheless, from our expertise with Cyber-Physical Systems, we are adamant that the architecture carries over to other such systems without major changes.
    \item \textbf{Starting state of virtual experiments:} For the gantry crane, the starting state for each simulation could be retrieved from the database since they were measured directly, that is, the position, velocity, angular position and angular velocity. However, this is not always the case; in more complex systems, such states may need to be estimated with techniques such as a particle filter or Kalman filter, which may impact the number of replications that can be computed due to an increase in computational load.
    \item \textbf{Computational load:} right now, the replicated simulations and data transfer to the database proved no issue regarding their computational load or bandwidth usage. If the system were scaled up to many more physical systems and twins, the computational load might start to pose an issue. This is an issue that should be tackled only when it is faced, for which common solutions such as database load balancing are available.
\end{itemize}

\subsection{Future: Continuous/Continual X}
\label{sec:future}

In software engineering, there already exists a set of ``continuous X'' techniques, such as continuous testing, integration, and deployment. These techniques are all located in the software development's design and deployment side. In the future, a welcome development would be that techniques such as the one proposed here, but also in \cite{Overbeck2021,Lugaresi2022,Lugaresi2023,Munoz2022} could be integrated at the operational side of the system, providing a continuous validation of the system. Perhaps some thought needs to be given to the adjective continuous, which has historically been used but has a connotation of meaning ``continuous without interruption''. In contrast, continual can be used for both ``continuous without interruption'' and ``occurring regularly''. At least in our case, ``occurring regularly'' is a correcter way of describing the validation process than ``continuous without interruption''.

%% file: text/g-conclusion.tex
\section{Conclusion}
\label{sec:conclusion}

This paper started from the problem statement that in works on digital twins, little thought is given to the quality assurance of the digital twin models. From a \ac{MandS} viewpoint, we noted that there has already been a lot of research into the quality assurance of models in the form of \ac{VandV} efforts during model development, though those techniques were usually not applied continuously during the system's operation. Therefore, we tested a set of metrics normally used during model validation, to check whether or not they are useful for evolving a digital twin alongside its real-world counterpart. The digital twin that we tested on was that of a miniaturized gantry crane, which differs from the related work, which so far mostly looked at production systems. Our results showed that, indeed, some of the metrics can be successfully used at runtime to detect changes in the real-world system and trigger a calibration of the digital twin. There do remain some caveats when applying the techniques, which we discussed in detail, but the general conclusion is that model validation techniques can be reused to support the evolution of twinned systems. Furthermore, we found the gantry crane setup proves a good testing ground for further digital twin research.

\section*{Declarations}
\subsection*{Funding}
Joost Mertens is funded by the Research Foundation - Flanders (FWO) through strategic basic research grant 1SD3421N.

%% file: text/h-appendix.tex
\begin{appendices}

\section{Database Tables}
\label{sec:database-tables}

\begin{figure}
    \centering
    \includegraphics[width=\textwidth]{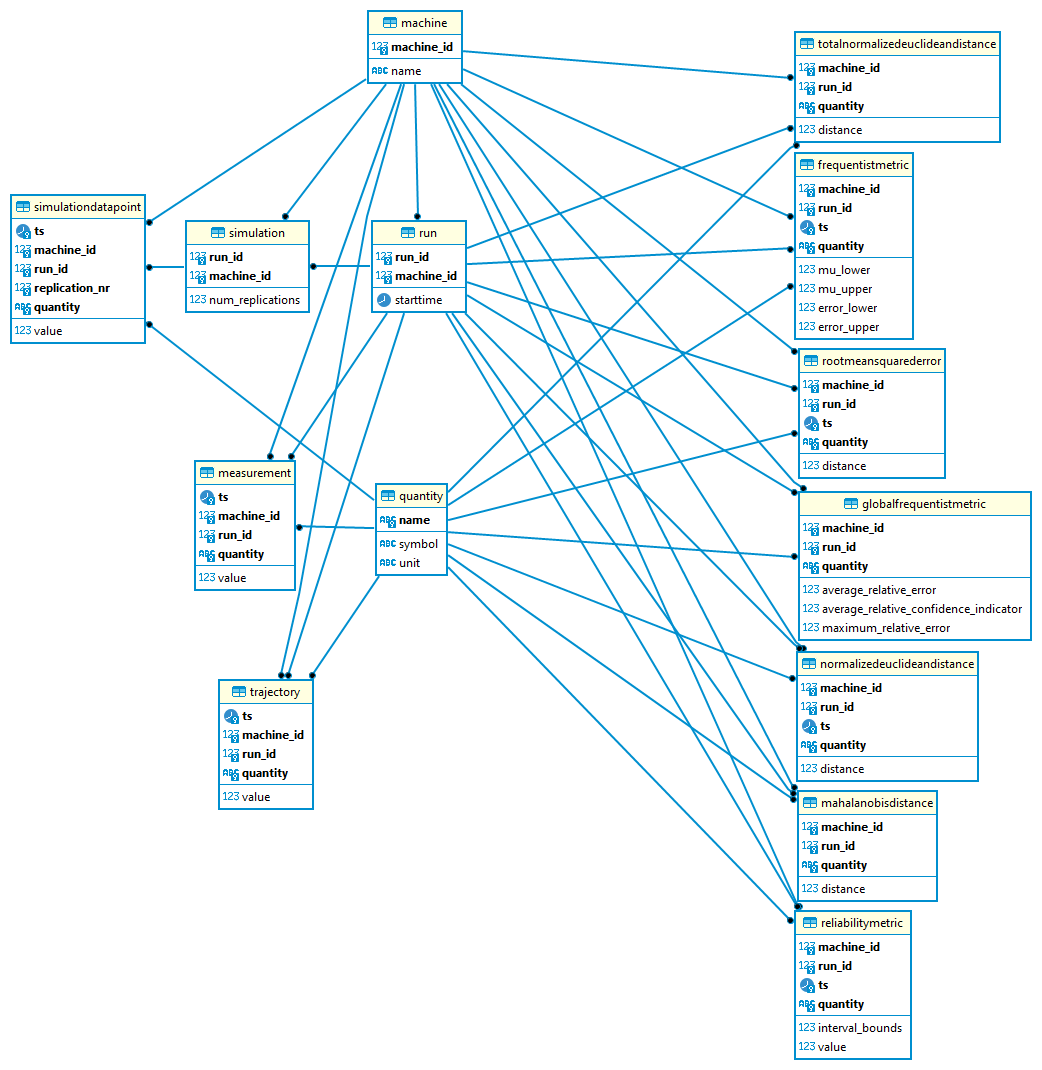}
    \caption{Entity Relation diagram of the database tables.}
    \label{fig:ER-diagram}
\end{figure}

\Cref{fig:ER-diagram} shows the entity relation diagram of the database. Unfortunately, the choice of composite primary keys makes the diagram a bit verbose. The diagram is defined with a machine's routine operation as vantage point. The choice was made to store the data in a narrow format. We'll discuss the tables next.

\begin{itemize}
    \item The \textbf{machine} table contains all the machines in the factory/workshop/laboratory. Each machine is given a unique ID and a name for easy human identification.
    \item The \textbf{quantity} table stores the quantities that can be used in the other tables. It can be used to get the unit or symbol of a quantity.
    \item The \textbf{run} table contains an entry for every routine operation. Each run is given a unique ID, is associated to a machine and holds a start time. The start time is used because it allows us to store all other data associated with this run in relative time from this starting time. This makes comparing the simulations, optimal trajectory and measurements with each other simpler.
    \item For each run, an optimal trajectory is calculated. Each of these trajectories is stored in the \textbf{trajectory} table. Every row of that table contains the value of a quantity for each time step (actually for each time step within a run associated with a specific machine, but conceptually it's for each time step, a similar remark holds for the \textbf{simulationdatapoint} and \textbf{measurement} tables).
    \item When a run is executed in the real world, measurements are made that are stored in the \textbf{measurement} table. Similarly to the \textbf{trajectory} table, each row stores the value of a quantity for a time step.
    \item Associated with each run there is a simulation, each of which has a number of replications. The simulations are given an entry in the \textbf{simulation} table, no unique ID is needed since a composite primary key of run ID and machine ID suffices. The entry states how many replications were simulated. For all those replications, simulation data points are collected, which are stored in the \textbf{simulationdatapoint} table, each row stores the value of a quantity for each time step and replication.
    \item Lastly, for each of the distance metrics, there exists a table to store their values. These are all on the right side of \cref{fig:ER-diagram}. The database shows one table for the Reliability metric, which did not end up in the results.
\end{itemize}

\end{appendices}

%% file: main.bbl

\begin{thebibliography}{48}
\ifx \bisbn   \undefined \def \bisbn  #1{ISBN #1}\fi
\ifx \binits  \undefined \def \binits#1{#1}\fi
\ifx \bauthor  \undefined \def \bauthor#1{#1}\fi
\ifx \batitle  \undefined \def \batitle#1{#1}\fi
\ifx \bjtitle  \undefined \def \bjtitle#1{#1}\fi
\ifx \bvolume  \undefined \def \bvolume#1{\textbf{#1}}\fi
\ifx \byear  \undefined \def \byear#1{#1}\fi
\ifx \bissue  \undefined \def \bissue#1{#1}\fi
\ifx \bfpage  \undefined \def \bfpage#1{#1}\fi
\ifx \blpage  \undefined \def \blpage #1{#1}\fi
\ifx \burl  \undefined \def \burl#1{\textsf{#1}}\fi
\ifx \doiurl  \undefined \def \doiurl#1{\url{https://doi.org/#1}}\fi
\ifx \betal  \undefined \def \betal{\textit{et al.}}\fi
\ifx \binstitute  \undefined \def \binstitute#1{#1}\fi
\ifx \binstitutionaled  \undefined \def \binstitutionaled#1{#1}\fi
\ifx \bctitle  \undefined \def \bctitle#1{#1}\fi
\ifx \beditor  \undefined \def \beditor#1{#1}\fi
\ifx \bpublisher  \undefined \def \bpublisher#1{#1}\fi
\ifx \bbtitle  \undefined \def \bbtitle#1{#1}\fi
\ifx \bedition  \undefined \def \bedition#1{#1}\fi
\ifx \bseriesno  \undefined \def \bseriesno#1{#1}\fi
\ifx \blocation  \undefined \def \blocation#1{#1}\fi
\ifx \bsertitle  \undefined \def \bsertitle#1{#1}\fi
\ifx \bsnm \undefined \def \bsnm#1{#1}\fi
\ifx \bsuffix \undefined \def \bsuffix#1{#1}\fi
\ifx \bparticle \undefined \def \bparticle#1{#1}\fi
\ifx \barticle \undefined \def \barticle#1{#1}\fi
\bibcommenthead
\ifx \bconfdate \undefined \def \bconfdate #1{#1}\fi
\ifx \botherref \undefined \def \botherref #1{#1}\fi
\ifx \url \undefined \def \url#1{\textsf{#1}}\fi
\ifx \bchapter \undefined \def \bchapter#1{#1}\fi
\ifx \bbook \undefined \def \bbook#1{#1}\fi
\ifx \bcomment \undefined \def \bcomment#1{#1}\fi
\ifx \oauthor \undefined \def \oauthor#1{#1}\fi
\ifx \citeauthoryear \undefined \def \citeauthoryear#1{#1}\fi
\ifx \endbibitem  \undefined \def \endbibitem {}\fi
\ifx \bconflocation  \undefined \def \bconflocation#1{#1}\fi
\ifx \arxivurl  \undefined \def \arxivurl#1{\textsf{#1}}\fi
\csname PreBibitemsHook\endcsname

\bibitem[\protect\citeauthoryear{Wright and Davidson}{2020}]{wright2020}
\begin{barticle}
\bauthor{\bsnm{Wright}, \binits{L.}},
\bauthor{\bsnm{Davidson}, \binits{S.}}:
\batitle{How to tell the difference between a model and a digital twin}.
\bjtitle{Advanced Modeling and Simulation in Engineering Sciences}
\bvolume{7}(\bissue{1}),
\bfpage{1}--\blpage{13}
(\byear{2020})
\doiurl{10.1186/s40323-020-00147-4}
\end{barticle}
\endbibitem

\bibitem[\protect\citeauthoryear{Dalibor et~al.}{2022}]{Dalibor2022}
\begin{barticle}
\bauthor{\bsnm{Dalibor}, \binits{M.}},
\bauthor{\bsnm{Jansen}, \binits{N.}},
\bauthor{\bsnm{Rumpe}, \binits{B.}},
\bauthor{\bsnm{Schmalzing}, \binits{D.}},
\bauthor{\bsnm{Wachtmeister}, \binits{L.}},
\bauthor{\bsnm{Wimmer}, \binits{M.}},
\bauthor{\bsnm{Wortmann}, \binits{A.}}:
\batitle{A cross-domain systematic mapping study on software engineering for
  digital twins}.
\bjtitle{Journal of Systems and Software}
\bvolume{193},
\bfpage{111361}
(\byear{2022})
\doiurl{10.1016/j.jss.2022.111361}
\end{barticle}
\endbibitem

\bibitem[\protect\citeauthoryear{Botín-Sanabria et~al.}{2022}]{botin2022}
\begin{botherref}
\oauthor{\bsnm{Botín-Sanabria}, \binits{D.M.}},
\oauthor{\bsnm{Mihaita}, \binits{A.-S.}},
\oauthor{\bsnm{Peimbert-García}, \binits{R.E.}},
\oauthor{\bsnm{Ramírez-Moreno}, \binits{M.A.}},
\oauthor{\bsnm{Ramírez-Mendoza}, \binits{R.A.}},
\oauthor{\bsnm{Lozoya-Santos}, \binits{J.d.J.}}:
Digital twin technology challenges and applications: A comprehensive review.
Remote Sensing
\textbf{14}(6)
(2022)
\doiurl{10.3390/rs14061335}
\end{botherref}
\endbibitem

\bibitem[\protect\citeauthoryear{Semeraro et~al.}{2021}]{semeraro2021}
\begin{barticle}
\bauthor{\bsnm{Semeraro}, \binits{C.}},
\bauthor{\bsnm{Lezoche}, \binits{M.}},
\bauthor{\bsnm{Panetto}, \binits{H.}},
\bauthor{\bsnm{Dassisti}, \binits{M.}}:
\batitle{Digital twin paradigm: A systematic literature review}.
\bjtitle{Computers in Industry}
\bvolume{130},
\bfpage{103469}
(\byear{2021})
\doiurl{10.1016/j.compind.2021.103469}
\end{barticle}
\endbibitem

\bibitem[\protect\citeauthoryear{Lee}{2008}]{Lee2008}
\begin{bchapter}
\bauthor{\bsnm{Lee}, \binits{E.A.}}:
\bctitle{Cyber physical systems: Design challenges}.
In: \bbtitle{2008 11th IEEE International Symposium on Object and
  Component-Oriented Real-Time Distributed Computing (ISORC)},
pp. \bfpage{363}--\blpage{369}
(\byear{2008}).
\doiurl{10.1109/ISORC.2008.25}
\end{bchapter}
\endbibitem

\bibitem[\protect\citeauthoryear{Sargent}{1984}]{Sargent1984}
\begin{bchapter}
\bauthor{\bsnm{Sargent}, \binits{R.G.}}:
\bctitle{A tutorial on verification and validation of simulation models}.
In: \bbtitle{Proceedings of the 16th Conference on Winter Simulation}.
\bsertitle{WSC '84},
pp. \bfpage{114}--\blpage{121}.
\bpublisher{IEEE Press},
\blocation{Piscataway, NJ, USA}
(\byear{1984})
\end{bchapter}
\endbibitem

\bibitem[\protect\citeauthoryear{Schlesinger et~al.}{1979}]{Schlesinger1979}
\begin{barticle}
\bauthor{\bsnm{Schlesinger}, \binits{S.}},
\bauthor{\bsnm{Crosbie}, \binits{R.E.}},
\bauthor{\bsnm{Gagné}, \binits{R.E.}},
\bauthor{\bsnm{Innis}, \binits{G.S.}},
\bauthor{\bsnm{Lalwani}, \binits{C.S.}},
\bauthor{\bsnm{Loch}, \binits{J.}},
\bauthor{\bsnm{Sylvester}, \binits{R.J.}},
\bauthor{\bsnm{Wright}, \binits{R.D.}},
\bauthor{\bsnm{Kheir}, \binits{N.}},
\bauthor{\bsnm{Bartos}, \binits{D.}}:
\batitle{Terminology for model credibility}.
\bjtitle{SIMULATION}
\bvolume{32}(\bissue{3}),
\bfpage{103}--\blpage{104}
(\byear{1979})
\doiurl{10.1177/003754977903200304}
\end{barticle}
\endbibitem

\bibitem[\protect\citeauthoryear{Balci}{2012}]{balci2012life}
\begin{barticle}
\bauthor{\bsnm{Balci}, \binits{O.}}:
\batitle{A life cycle for modeling and simulation}.
\bjtitle{Simulation}
\bvolume{88}(\bissue{7}),
\bfpage{870}--\blpage{883}
(\byear{2012})
\end{barticle}
\endbibitem

\bibitem[\protect\citeauthoryear{Singh et~al.}{2021}]{Singh2021}
\begin{botherref}
\oauthor{\bsnm{Singh}, \binits{M.}},
\oauthor{\bsnm{Fuenmayor}, \binits{E.}},
\oauthor{\bsnm{Hinchy}, \binits{E.P.}},
\oauthor{\bsnm{Qiao}, \binits{Y.}},
\oauthor{\bsnm{Murray}, \binits{N.}},
\oauthor{\bsnm{Devine}, \binits{D.}}:
Digital twin: Origin to future.
Applied System Innovation
\textbf{4}(2)
(2021)
\doiurl{10.3390/asi4020036}
\end{botherref}
\endbibitem

\bibitem[\protect\citeauthoryear{{International Council on Systems Engineering
  (INCOSE)}}{2021}]{INCOSE2021}
\begin{botherref}
\oauthor{\bsnm{{International Council on Systems Engineering (INCOSE)}}}:
Incose systems engineering vision 2035.
Technical report,
{International Council on Systems Engineering (INCOSE)}
(2021)
\end{botherref}
\endbibitem

\bibitem[\protect\citeauthoryear{Modoni et~al.}{2019}]{Modoni2019}
\begin{barticle}
\bauthor{\bsnm{Modoni}, \binits{G.E.}},
\bauthor{\bsnm{Caldarola}, \binits{E.G.}},
\bauthor{\bsnm{Sacco}, \binits{M.}},
\bauthor{\bsnm{Terkaj}, \binits{W.}}:
\batitle{Synchronizing physical and digital factory: benefits and technical
  challenges}.
\bjtitle{Procedia CIRP}
\bvolume{79},
\bfpage{472}--\blpage{477}
(\byear{2019})
\doiurl{10.1016/j.procir.2019.02.125} .
\bcomment{12th CIRP Conference on Intelligent Computation in Manufacturing
  Engineering, 18-20 July 2018, Gulf of Naples, Italy}
\end{barticle}
\endbibitem

\bibitem[\protect\citeauthoryear{Glaessgen and Stargel}{2012}]{Glaessgen2012}
\begin{bbook}
\bauthor{\bsnm{Glaessgen}, \binits{E.}},
\bauthor{\bsnm{Stargel}, \binits{D.}}:
\bbtitle{The Digital Twin Paradigm for Future NASA and U.S. Air Force
  Vehicles},
(\byear{2012}).
\doiurl{10.2514/6.2012-1818}
\end{bbook}
\endbibitem

\bibitem[\protect\citeauthoryear{Grieves and Vickers}{2017}]{Grieves2017}
\begin{bbook}
\bauthor{\bsnm{Grieves}, \binits{M.}},
\bauthor{\bsnm{Vickers}, \binits{J.}}:
In: \beditor{\bsnm{Kahlen}, \binits{F.-J.}},
\beditor{\bsnm{Flumerfelt}, \binits{S.}},
\beditor{\bsnm{Alves}, \binits{A.}} (eds.)
\bbtitle{Digital Twin: Mitigating Unpredictable, Undesirable Emergent Behavior
  in Complex Systems},
pp. \bfpage{85}--\blpage{113}.
\bpublisher{Springer},
\blocation{Cham}
(\byear{2017}).
\doiurl{10.1007/978-3-319-38756-7_4}
\end{bbook}
\endbibitem

\bibitem[\protect\citeauthoryear{Council}{2012}]{NAP2012}
\begin{bbook}
\bauthor{\bsnm{Council}, \binits{N.R.}}:
\bbtitle{Assessing the Reliability of Complex Models: Mathematical and
  Statistical Foundations of Verification, Validation, and Uncertainty
  Quantification}.
\bpublisher{The National Academies Press},
\blocation{Washington, DC}
(\byear{2012}).
\doiurl{10.17226/13395}
\end{bbook}
\endbibitem

\bibitem[\protect\citeauthoryear{Fishman and Kiviat}{1968}]{Fishman1968}
\begin{barticle}
\bauthor{\bsnm{Fishman}, \binits{G.S.}},
\bauthor{\bsnm{Kiviat}, \binits{P.J.}}:
\batitle{The statistics of discrete-event simulation}.
\bjtitle{SIMULATION}
\bvolume{10}(\bissue{4}),
\bfpage{185}--\blpage{195}
(\byear{1968})
\doiurl{10.1177/003754976801000406}
\end{barticle}
\endbibitem

\bibitem[\protect\citeauthoryear{Oberkampf and Roy}{2010}]{Oberkampf2010}
\begin{bbook}
\bauthor{\bsnm{Oberkampf}, \binits{W.L.}},
\bauthor{\bsnm{Roy}, \binits{C.J.}}:
\bbtitle{Verification and Validation in Scientific Computing}.
\bpublisher{Cambridge University Press},
\blocation{Cambridge}
(\byear{2010}).
\doiurl{10.1017/CBO9780511760396}
\end{bbook}
\endbibitem

\bibitem[\protect\citeauthoryear{Law}{2014}]{law2014simulation}
\begin{bbook}
\bauthor{\bsnm{Law}, \binits{A.M.}}:
\bbtitle{Simulation Modeling and Analysis}.
\bpublisher{McGraw-Hill Education},
\blocation{New York}
(\byear{2014})
\end{bbook}
\endbibitem

\bibitem[\protect\citeauthoryear{Beisbart and
  Saam}{2019}]{beisbart2019computer}
\begin{bbook}
\bauthor{\bsnm{Beisbart}, \binits{C.}},
\bauthor{\bsnm{Saam}, \binits{N.J.}}:
\bbtitle{Computer Simulation Validation}.
\bpublisher{Springer},
\blocation{Switzerland}
(\byear{2019}).
\doiurl{10.1007/978-3-319-70766-2}
\end{bbook}
\endbibitem

\bibitem[\protect\citeauthoryear{Maupin and Swiller}{2017}]{Maupin2017}
\begin{botherref}
\oauthor{\bsnm{Maupin}, \binits{K.A.}},
\oauthor{\bsnm{Swiller}, \binits{L.P.}}:
Validation metrics for deterministic and probabilistic data.
Technical report,
Sandia National Laboratories
(2017)
\end{botherref}
\endbibitem

\bibitem[\protect\citeauthoryear{Vanslette et~al.}{2020}]{Vanslette2019}
\begin{barticle}
\bauthor{\bsnm{Vanslette}, \binits{K.}},
\bauthor{\bsnm{Tohme}, \binits{T.}},
\bauthor{\bsnm{Youcef-Toumi}, \binits{K.}}:
\batitle{A general model validation and testing tool}.
\bjtitle{Reliability Engineering \& System Safety}
\bvolume{195},
\bfpage{106684}
(\byear{2020})
\doiurl{10.1016/j.ress.2019.106684}
\end{barticle}
\endbibitem

\bibitem[\protect\citeauthoryear{Ferson et~al.}{2008}]{Ferson2008}
\begin{barticle}
\bauthor{\bsnm{Ferson}, \binits{S.}},
\bauthor{\bsnm{Oberkampf}, \binits{W.L.}},
\bauthor{\bsnm{Ginzburg}, \binits{L.}}:
\batitle{Model validation and predictive capability for the thermal challenge
  problem}.
\bjtitle{Computer Methods in Applied Mechanics and Engineering}
\bvolume{197}(\bissue{29}),
\bfpage{2408}--\blpage{2430}
(\byear{2008})
\doiurl{10.1016/j.cma.2007.07.030} .
\bcomment{Validation Challenge Workshop}
\end{barticle}
\endbibitem

\bibitem[\protect\citeauthoryear{Roy and Oberkampf}{2011}]{Roy2011}
\begin{barticle}
\bauthor{\bsnm{Roy}, \binits{C.J.}},
\bauthor{\bsnm{Oberkampf}, \binits{W.L.}}:
\batitle{A comprehensive framework for verification, validation, and
  uncertainty quantification in scientific computing}.
\bjtitle{Computer Methods in Applied Mechanics and Engineering}
\bvolume{200}(\bissue{25}),
\bfpage{2131}--\blpage{2144}
(\byear{2011})
\doiurl{10.1016/j.cma.2011.03.016}
\end{barticle}
\endbibitem

\bibitem[\protect\citeauthoryear{Rebba and Mahadevan}{2008}]{Rebba2008}
\begin{barticle}
\bauthor{\bsnm{Rebba}, \binits{R.}},
\bauthor{\bsnm{Mahadevan}, \binits{S.}}:
\batitle{Computational methods for model reliability assessment}.
\bjtitle{Reliability Engineering \& System Safety}
\bvolume{93}(\bissue{8}),
\bfpage{1197}--\blpage{1207}
(\byear{2008})
\doiurl{10.1016/j.ress.2007.08.001}
\end{barticle}
\endbibitem

\bibitem[\protect\citeauthoryear{Oberkampf and Barone}{2006}]{Oberkampf2006}
\begin{barticle}
\bauthor{\bsnm{Oberkampf}, \binits{W.L.}},
\bauthor{\bsnm{Barone}, \binits{M.F.}}:
\batitle{Measures of agreement between computation and experiment: Validation
  metrics}.
\bjtitle{Journal of Computational Physics}
\bvolume{217}(\bissue{1}),
\bfpage{5}--\blpage{36}
(\byear{2006})
\doiurl{10.1016/j.jcp.2006.03.037}
\end{barticle}
\endbibitem

\bibitem[\protect\citeauthoryear{Overbeck et~al.}{2021}]{Overbeck2021}
\begin{bchapter}
\bauthor{\bsnm{Overbeck}, \binits{L.}},
\bauthor{\bsnm{{Le Louarn}}, \binits{A.}},
\bauthor{\bsnm{Brützel}, \binits{O.}},
\bauthor{\bsnm{Stricker}, \binits{N.}},
\bauthor{\bsnm{Lanza}, \binits{G.}}:
\bctitle{Continuous validation and updating for high accuracy of digital twins
  of production systems}.
In: \bbtitle{Simulation in Produktion und Logistik 2021, Erlangen,
  15.-17.September 2021. Hrsg.: J. Franke}.
\bsertitle{Simulation in Produktion und Logistik 2021},
vol. \bseriesno{0},
pp. \bfpage{609}--\blpage{617}.
\bpublisher{{Cuvillier Verlag}},
\blocation{Göttingen, Germany}
(\byear{2021})
\end{bchapter}
\endbibitem

\bibitem[\protect\citeauthoryear{Lugaresi et~al.}{2022}]{Lugaresi2022}
\begin{bchapter}
\bauthor{\bsnm{Lugaresi}, \binits{G.}},
\bauthor{\bsnm{Gangemi}, \binits{S.}},
\bauthor{\bsnm{Gazzoni}, \binits{G.}},
\bauthor{\bsnm{Matta}, \binits{A.}}:
\bctitle{Online validation of simulation-based digital twins exploiting time
  series analysis}.
In: \bbtitle{2022 Winter Simulation Conference (WSC)},
pp. \bfpage{2912}--\blpage{2923}
(\byear{2022}).
\doiurl{10.1109/WSC57314.2022.10015346}
\end{bchapter}
\endbibitem

\bibitem[\protect\citeauthoryear{Lugaresi et~al.}{2023}]{Lugaresi2023}
\begin{barticle}
\bauthor{\bsnm{Lugaresi}, \binits{G.}},
\bauthor{\bsnm{Gangemi}, \binits{S.}},
\bauthor{\bsnm{Gazzoni}, \binits{G.}},
\bauthor{\bsnm{Matta}, \binits{A.}}:
\batitle{Online validation of digital twins for manufacturing systems}.
\bjtitle{Computers in Industry}
\bvolume{150},
\bfpage{103942}
(\byear{2023})
\doiurl{10.1016/j.compind.2023.103942}
\end{barticle}
\endbibitem

\bibitem[\protect\citeauthoryear{Mu\~{n}oz et~al.}{2022}]{Munoz2022}
\begin{bchapter}
\bauthor{\bsnm{Mu\~{n}oz}, \binits{P.}},
\bauthor{\bsnm{Wimmer}, \binits{M.}},
\bauthor{\bsnm{Troya}, \binits{J.}},
\bauthor{\bsnm{Vallecillo}, \binits{A.}}:
\bctitle{Using trace alignments for measuring the similarity between a physical
  and its digital twin}.
In: \bbtitle{Proceedings of the 25th International Conference on Model Driven
  Engineering Languages and Systems: Companion Proceedings}.
\bsertitle{MODELS '22},
pp. \bfpage{503}--\blpage{510}.
\bpublisher{Association for Computing Machinery},
\blocation{New York, NY, USA}
(\byear{2022}).
\doiurl{10.1145/3550356.3563135}
\end{bchapter}
\endbibitem

\bibitem[\protect\citeauthoryear{Gertler}{1998}]{gertler1998}
\begin{bbook}
\bauthor{\bsnm{Gertler}, \binits{J.}}:
\bbtitle{Fault Detection and Diagnosis in Engineering Systems}.
\bpublisher{CRC press},
\blocation{Boca Raton}
(\byear{1998})
\end{bbook}
\endbibitem

\bibitem[\protect\citeauthoryear{Schmidt et~al.}{2020}]{Schmidt2020}
\begin{bchapter}
\bauthor{\bsnm{Schmidt}, \binits{T.}},
\bauthor{\bsnm{Hauer}, \binits{F.}},
\bauthor{\bsnm{Pretschner}, \binits{A.}}:
\bctitle{Automated anomaly detection in cps log files}.
In: \beditor{\bsnm{Casimiro}, \binits{A.}},
\beditor{\bsnm{Ortmeier}, \binits{F.}},
\beditor{\bsnm{Bitsch}, \binits{F.}},
\beditor{\bsnm{Ferreira}, \binits{P.}} (eds.)
\bbtitle{Computer Safety, Reliability, and Security},
pp. \bfpage{179}--\blpage{194}.
\bpublisher{Springer},
\blocation{Cham}
(\byear{2020})
\end{bchapter}
\endbibitem

\bibitem[\protect\citeauthoryear{Harada et~al.}{2017}]{Harada2017}
\begin{bchapter}
\bauthor{\bsnm{Harada}, \binits{Y.}},
\bauthor{\bsnm{Yamagata}, \binits{Y.}},
\bauthor{\bsnm{Mizuno}, \binits{O.}},
\bauthor{\bsnm{Choi}, \binits{E.-H.}}:
\bctitle{Log-based anomaly detection of cps using a statistical method}.
In: \bbtitle{2017 8th International Workshop on Empirical Software Engineering
  in Practice (IWESEP)},
pp. \bfpage{1}--\blpage{6}
(\byear{2017}).
\doiurl{10.1109/IWESEP.2017.12}
\end{bchapter}
\endbibitem

\bibitem[\protect\citeauthoryear{Narasimhan and Biswas}{2007}]{Narasimhan2007}
\begin{barticle}
\bauthor{\bsnm{Narasimhan}, \binits{S.}},
\bauthor{\bsnm{Biswas}, \binits{G.}}:
\batitle{Model-based diagnosis of hybrid systems}.
\bjtitle{IEEE Transactions on Systems, Man, and Cybernetics - Part A: Systems
  and Humans}
\bvolume{37}(\bissue{3}),
\bfpage{348}--\blpage{361}
(\byear{2007})
\doiurl{10.1109/TSMCA.2007.893487}
\end{barticle}
\endbibitem

\bibitem[\protect\citeauthoryear{Zhao et~al.}{2005}]{Zhao2005}
\begin{barticle}
\bauthor{\bsnm{Zhao}, \binits{F.}},
\bauthor{\bsnm{Koutsoukos}, \binits{X.}},
\bauthor{\bsnm{Haussecker}, \binits{H.}},
\bauthor{\bsnm{Reich}, \binits{J.}},
\bauthor{\bsnm{Cheung}, \binits{P.}}:
\batitle{Monitoring and fault diagnosis of hybrid systems}.
\bjtitle{IEEE Transactions on Systems, Man, and Cybernetics, Part B
  (Cybernetics)}
\bvolume{35}(\bissue{6}),
\bfpage{1225}--\blpage{1240}
(\byear{2005})
\doiurl{10.1109/TSMCB.2005.850178}
\end{barticle}
\endbibitem

\bibitem[\protect\citeauthoryear{Feng et~al.}{2021}]{Feng2021}
\begin{bchapter}
\bauthor{\bsnm{Feng}, \binits{H.}},
\bauthor{\bsnm{Gomes}, \binits{C.}},
\bauthor{\bsnm{Thule}, \binits{C.}},
\bauthor{\bsnm{Lausdahl}, \binits{K.}},
\bauthor{\bsnm{Iosifidis}, \binits{A.}},
\bauthor{\bsnm{Larsen}, \binits{P.G.}}:
\bctitle{Introduction to digital twin engineering}.
In: \bbtitle{2021 Annual Modeling and Simulation Conference (ANNSIM)},
pp. \bfpage{1}--\blpage{12}
(\byear{2021}).
\doiurl{10.23919/ANNSIM52504.2021.9552135}
\end{bchapter}
\endbibitem

\bibitem[\protect\citeauthoryear{Luo et~al.}{2021}]{Luo2021}
\begin{botherref}
\oauthor{\bsnm{Luo}, \binits{Y.}},
\oauthor{\bsnm{Xiao}, \binits{Y.}},
\oauthor{\bsnm{Cheng}, \binits{L.}},
\oauthor{\bsnm{Peng}, \binits{G.}},
\oauthor{\bsnm{Yao}, \binits{D.D.}}:
Deep learning-based anomaly detection in cyber-physical systems: Progress and
  opportunities.
ACM Comput. Surv.
\textbf{54}(5)
(2021)
\doiurl{10.1145/3453155}
\end{botherref}
\endbibitem

\bibitem[\protect\citeauthoryear{Xu et~al.}{2021}]{Xu2021}
\begin{bchapter}
\bauthor{\bsnm{Xu}, \binits{Q.}},
\bauthor{\bsnm{Ali}, \binits{S.}},
\bauthor{\bsnm{Yue}, \binits{T.}}:
\bctitle{Digital twin-based anomaly detection in cyber-physical systems}.
In: \bbtitle{2021 14th IEEE Conference on Software Testing, Verification and
  Validation (ICST)},
pp. \bfpage{205}--\blpage{216}
(\byear{2021}).
\doiurl{10.1109/ICST49551.2021.00031}
\end{bchapter}
\endbibitem

\bibitem[\protect\citeauthoryear{Kritzinger et~al.}{2018}]{Kritzinger2018}
\begin{barticle}
\bauthor{\bsnm{Kritzinger}, \binits{W.}},
\bauthor{\bsnm{Karner}, \binits{M.}},
\bauthor{\bsnm{Traar}, \binits{G.}},
\bauthor{\bsnm{Henjes}, \binits{J.}},
\bauthor{\bsnm{Sihn}, \binits{W.}}:
\batitle{Digital twin in manufacturing: A categorical literature review and
  classification}.
\bjtitle{IFAC-PapersOnLine}
\bvolume{51}(\bissue{11}),
\bfpage{1016}--\blpage{1022}
(\byear{2018})
\doiurl{10.1016/j.ifacol.2018.08.474} .
\bcomment{16th IFAC Symposium on Information Control Problems in Manufacturing
  INCOM 2018}
\end{barticle}
\endbibitem

\bibitem[\protect\citeauthoryear{Mertens and Denil}{2021}]{Mertens2021}
\begin{bchapter}
\bauthor{\bsnm{Mertens}, \binits{J.}},
\bauthor{\bsnm{Denil}, \binits{J.}}:
\bctitle{The digital twin as a common knowledge base in devops to support
  continuous system evolution}.
In: \beditor{\bsnm{Habli}, \binits{I.}},
\beditor{\bsnm{Sujan}, \binits{M.}},
\beditor{\bsnm{Gerasimou}, \binits{S.}},
\beditor{\bsnm{Schoitsch}, \binits{E.}},
\beditor{\bsnm{Bitsch}, \binits{F.}} (eds.)
\bbtitle{Computer Safety, Reliability, and Security. SAFECOMP 2021 Workshops},
pp. \bfpage{158}--\blpage{170}.
\bpublisher{Springer},
\blocation{Cham}
(\byear{2021})
\end{bchapter}
\endbibitem

\bibitem[\protect\citeauthoryear{Mertens and Denil}{2023}]{Mertens2023}
\begin{bchapter}
\bauthor{\bsnm{Mertens}, \binits{J.}},
\bauthor{\bsnm{Denil}, \binits{J.}}:
\bctitle{Digital-twin co-evolution using continuous validation}.
In: \bbtitle{Proceedings of the ACM/IEEE 14th International Conference on
  Cyber-Physical Systems (with CPS-IoT Week 2023)}.
\bsertitle{ICCPS '23},
pp. \bfpage{266}--\blpage{267}.
\bpublisher{Association for Computing Machinery},
\blocation{New York, NY, USA}
(\byear{2023}).
\doiurl{10.1145/3576841.3589628}
\end{bchapter}
\endbibitem

\bibitem[\protect\citeauthoryear{Virtanen et~al.}{2020}]{2020SciPy}
\begin{barticle}
\bauthor{\bsnm{Virtanen}, \binits{P.}},
\bauthor{\bsnm{Gommers}, \binits{R.}},
\bauthor{\bsnm{Oliphant}, \binits{T.E.}},
\bauthor{\bsnm{Haberland}, \binits{M.}},
\bauthor{\bsnm{Reddy}, \binits{T.}},
\bauthor{\bsnm{Cournapeau}, \binits{D.}},
\bauthor{\bsnm{Burovski}, \binits{E.}},
\bauthor{\bsnm{Peterson}, \binits{P.}},
\bauthor{\bsnm{Weckesser}, \binits{W.}},
\bauthor{\bsnm{Bright}, \binits{J.}},
\bauthor{\bsnm{{van der Walt}}, \binits{S.J.}},
\bauthor{\bsnm{Brett}, \binits{M.}},
\bauthor{\bsnm{Wilson}, \binits{J.}},
\bauthor{\bsnm{Millman}, \binits{K.J.}},
\bauthor{\bsnm{Mayorov}, \binits{N.}},
\bauthor{\bsnm{Nelson}, \binits{A.R.J.}},
\bauthor{\bsnm{Jones}, \binits{E.}},
\bauthor{\bsnm{Kern}, \binits{R.}},
\bauthor{\bsnm{Larson}, \binits{E.}},
\bauthor{\bsnm{Carey}, \binits{C.J.}},
\bauthor{\bsnm{Polat}, \binits{{\. I}.}},
\bauthor{\bsnm{Feng}, \binits{Y.}},
\bauthor{\bsnm{Moore}, \binits{E.W.}},
\bauthor{\bsnm{{VanderPlas}}, \binits{J.}},
\bauthor{\bsnm{Laxalde}, \binits{D.}},
\bauthor{\bsnm{Perktold}, \binits{J.}},
\bauthor{\bsnm{Cimrman}, \binits{R.}},
\bauthor{\bsnm{Henriksen}, \binits{I.}},
\bauthor{\bsnm{Quintero}, \binits{E.A.}},
\bauthor{\bsnm{Harris}, \binits{C.R.}},
\bauthor{\bsnm{Archibald}, \binits{A.M.}},
\bauthor{\bsnm{Ribeiro}, \binits{A.H.}},
\bauthor{\bsnm{Pedregosa}, \binits{F.}},
\bauthor{\bsnm{{van Mulbregt}}, \binits{P.}},
\bauthor{\bsnm{{SciPy 1.0 Contributors}}}:
\batitle{{{SciPy} 1.0: Fundamental Algorithms for Scientific Computing in
  Python}}.
\bjtitle{Nature Methods}
\bvolume{17},
\bfpage{261}--\blpage{272}
(\byear{2020})
\doiurl{10.1038/s41592-019-0686-2}
\end{barticle}
\endbibitem

\bibitem[\protect\citeauthoryear{}{}]{matlabparameterestimator}
\begin{botherref}
Estimate model parameters and initial states.
\url{https://nl.mathworks.com/help/sldo/ref/parameterestimator-app.html}.
Accessed: 2024-05-03
\end{botherref}
\endbibitem

\bibitem[\protect\citeauthoryear{Schälte et~al.}{2023}]{schalte2023}
\begin{barticle}
\bauthor{\bsnm{Schälte}, \binits{Y.}},
\bauthor{\bsnm{Fröhlich}, \binits{F.}},
\bauthor{\bsnm{Jost}, \binits{P.J.}},
\bauthor{\bsnm{Vanhoefer}, \binits{J.}},
\bauthor{\bsnm{Pathirana}, \binits{D.}},
\bauthor{\bsnm{Stapor}, \binits{P.}},
\bauthor{\bsnm{Lakrisenko}, \binits{P.}},
\bauthor{\bsnm{Wang}, \binits{D.}},
\bauthor{\bsnm{Raimúndez}, \binits{E.}},
\bauthor{\bsnm{Merkt}, \binits{S.}},
\bauthor{\bsnm{Schmiester}, \binits{L.}},
\bauthor{\bsnm{Städter}, \binits{P.}},
\bauthor{\bsnm{Grein}, \binits{S.}},
\bauthor{\bsnm{Dudkin}, \binits{E.}},
\bauthor{\bsnm{Doresic}, \binits{D.}},
\bauthor{\bsnm{Weindl}, \binits{D.}},
\bauthor{\bsnm{Hasenauer}, \binits{J.}}:
\batitle{{pyPESTO: a modular and scalable tool for parameter estimation for
  dynamic models}}.
\bjtitle{Bioinformatics}
\bvolume{39}(\bissue{11}),
\bfpage{711}
(\byear{2023})
\doiurl{10.1093/bioinformatics/btad711}
{\href{https://arxiv.org/abs/https://academic.oup.com/bioinformatics/article-pdf/39/11/btad711/53962204/btad711.pdf}{{https://academic.oup.com/bioinformatics/article-pdf/39/11/btad711/53962204/btad711.pdf}}}
\end{barticle}
\endbibitem

\bibitem[\protect\citeauthoryear{Lauer and Bloch}{2019}]{lauer2019}
\begin{bbook}
\bauthor{\bsnm{Lauer}, \binits{F.}},
\bauthor{\bsnm{Bloch}, \binits{G.}}:
\bbtitle{Hybrid System Identification}.
\bpublisher{Springer}, \blocation{???}
(\byear{2019})
\end{bbook}
\endbibitem

\bibitem[\protect\citeauthoryear{Moradvandi et~al.}{2023}]{MORADVANDI2023}
\begin{barticle}
\bauthor{\bsnm{Moradvandi}, \binits{A.}},
\bauthor{\bsnm{Lindeboom}, \binits{R.E.F.}},
\bauthor{\bsnm{Abraham}, \binits{E.}},
\bauthor{\bsnm{{De Schutter}}, \binits{B.}}:
\batitle{Models and methods for hybrid system identification: a systematic
  survey*}.
\bjtitle{IFAC-PapersOnLine}
\bvolume{56}(\bissue{2}),
\bfpage{95}--\blpage{107}
(\byear{2023})
\doiurl{10.1016/j.ifacol.2023.10.1553} .
\bcomment{22nd IFAC World Congress}
\end{barticle}
\endbibitem

\bibitem[\protect\citeauthoryear{Liu et~al.}{2011}]{liu2011}
\begin{barticle}
\bauthor{\bsnm{Liu}, \binits{Y.}},
\bauthor{\bsnm{Chen}, \binits{W.}},
\bauthor{\bsnm{Arendt}, \binits{P.}},
\bauthor{\bsnm{Huang}, \binits{H.-Z.}}:
\batitle{{Toward a Better Understanding of Model Validation Metrics}}.
\bjtitle{Journal of Mechanical Design}
\bvolume{133}(\bissue{7}),
\bfpage{071005}
(\byear{2011})
\doiurl{10.1115/1.4004223}
{\href{https://arxiv.org/abs/https://asmedigitalcollection.asme.org/mechanicaldesign/article-pdf/133/7/071005/5926232/071005\_1.pdf}{{https://asmedigitalcollection.asme.org/mechanicaldesign/article-pdf/133/7/071005/5926232/071005\_1.pdf}}}
\end{barticle}
\endbibitem

\bibitem[\protect\citeauthoryear{Gillis et~al.}{2020}]{gillis2020}
\begin{bchapter}
\bauthor{\bsnm{Gillis}, \binits{J.}},
\bauthor{\bsnm{Vandewal}, \binits{B.}},
\bauthor{\bsnm{Pipeleers}, \binits{G.}},
\bauthor{\bsnm{Swevers}, \binits{J.}}:
\bctitle{Effortless modeling of optimal control problems with rockit}.
In: \bbtitle{39th Benelux Meeting on Systems and Control},
vol. \bseriesno{138}
(\byear{2020}).
\bcomment{The Netherlands}
\end{bchapter}
\endbibitem

\bibitem[\protect\citeauthoryear{Light}{2017}]{Light2017}
\begin{barticle}
\bauthor{\bsnm{Light}, \binits{R.A.}}:
\batitle{Mosquitto: server and client implementation of the {MQTT} protocol}.
\bjtitle{Journal of Open Source Software}
\bvolume{2}(\bissue{13}),
\bfpage{265}
(\byear{2017})
\doiurl{10.21105/joss.00265}
\end{barticle}
\endbibitem

\bibitem[\protect\citeauthoryear{Grieves}{2012}]{Grieves2012}
\begin{bchapter}
\bauthor{\bsnm{Grieves}, \binits{M.W.}}:
\bctitle{Virtually indistinguishable}.
In: \beditor{\bsnm{Rivest}, \binits{L.}},
\beditor{\bsnm{Bouras}, \binits{A.}},
\beditor{\bsnm{Louhichi}, \binits{B.}} (eds.)
\bbtitle{Product Lifecycle Management. Towards Knowledge-Rich Enterprises},
pp. \bfpage{226}--\blpage{242}.
\bpublisher{Springer},
\blocation{Berlin, Heidelberg}
(\byear{2012})
\end{bchapter}
\endbibitem

\end{thebibliography}
